\RequirePackage{fix-cm}
\documentclass[smallextended]{svjour3}       
\smartqed  
\usepackage{graphicx}
\usepackage{xcolor}
\usepackage[bookmarks=false]{hyperref}
\usepackage[numbers]{natbib}
\usepackage{longtable}
\usepackage{paralist}
\usepackage[justification=centering]{caption}
\usepackage{subfigure}

\begin{document}

\title{Artefact-based Requirements Engineering: \\
The AMDiRE Approach
}

\titlerunning{The AMDiRE Approach}        

\author{D. M\'{e}ndez Fern\'{a}ndez \and
        B. Penzenstadler 
}

\institute{Daniel M\'{e}ndez Fern\'{a}ndez \at
              Technische Universit\"at M\"unchen, Garching\\
              Tel.: +49-89-28917056\\
              \email{mendezfe@in.tum.de}           
           \and
           Birgit Penzenstadler \at
           University of California, Irvine\\
           \email{bpenzens@uci.edu}
}

\date{Received: date / Accepted: date}

\maketitle

\begin{abstract}
The various influences in the processes and application domains make Requirements Engineering (RE) inherently complex and difficult to implement. In general, we have two options for establishing an RE approach: we can either establish an activity-based RE approach or we can establish an artefact-based one where project participants concentrate on the RE artefacts rather than on the way of creating them. While a number of activity-based RE approaches have been proposed in recent years, we have gained much empirical evidence and experiences about the advantages of the artefact-based paradigm for RE. However, artefact orientation is still a young paradigm with various interpretations and practical manifestations whereby we need a clear understanding of its basic concepts and a consolidated and evaluated view on the paradigm. 

In this article, we contribute an artefact-based approach to RE (AMDiRE) that emerges from six years of experiences in fundamental and evidence-based research. To this end, we first discuss the basic notion of artefact orientation and its evolution in recent years. We briefly introduce a set of artefact-based RE models we developed in industrial research cooperations for different application domains, show their empirical evaluations, and their dissemination into academia and practice, eventually leading to the AMDiRE approach. We conclude with a discussion of experiences we made during the development and different industrial evaluations, and lessons learnt.

\keywords{Requirements Engineering \and Artefact Orientation \and Empirical Evaluation}
\end{abstract}

\section{Introduction}
\label{sec:Intro}

Requirements Engineering (RE) is an important success factor for software and systems development projects as precise requirements are critical determinants of quality~\cite{broy06_mbRE}. Although the discipline is known to be crucial for the success of every project, we still observe companies struggling with their RE process. Many of these companies have unclear roles and responsibilities but a detailedly defined process that is obligatory for all projects. RE is too often performed mindlessly or even faked~\cite{PC86}, without awareness of the reasons why a process step should (or should not) be executed, and without awareness of how to structure and specify the results~\cite{MWb13}. 

A major reason for this circumstance is that many things are not clear from the beginning of a project, which makes the discipline inherently complex and volatile. The need for flexibility is additionally hardened by potentially large amounts of requirements~\cite{RBW08}, which are too often insufficiently structured in spreadsheets. The effects of this circumstance can be often observed in incomplete and inconsistent requirements, and, finally, in failed projects.

The \emph{chaos report} from the Standish Group~\cite{chaos1995} states that 44~\% of the reasons for failed projects have their origin in insufficient RE. As the report takes only a limited view into RE itself and is also known to have serious flaws in its design negatively affecting the validity of the results~\cite{EV10}, we launched a series of empirical investigations on practical problems in RE and how those problems manifest themselves in the whole software development process~\cite{MW13, MW13b}. We discovered that the missing awareness of what should be done in RE manifests in irreproducible, incomplete, and inconsistent artefacts without clear terminology, all together seen to be the major reason for time overruns, cost overruns, and eventually for failed projects~\cite{MW13, MWLBC10}. A solution to these problems is to establish a company-wide \emph{RE reference model} that should support
\begin{compactenum}
\item flexibility in the way of working to cope with the various influences in individual project environments, and
\item the reproducible creation of resilient and detailed specification documents.
\end{compactenum}

In Zave's classification of research efforts in RE, this addresses the two problems of integrating multiple views and representations, and obtaining complete, consistent, and unambiguous specifications~\cite[p.317]{zave97}.

There are two basic paradigms for the establishment of such an RE reference model: activity orientation and artefact orientation. Activity orientation means to define the reference model by means of detailed interconnected procedures that dictate which methods to combine and use in which project situation~\cite{TB12}. The underlying idea is to define a situation-specific process by a set of small steps, i.e.\ methods to be performed in a particular order to create certain artefacts as outcome (see also~\cite{Brink96, AHV97}). In contrast, artefact orientation establishes a blueprint of the created RE results, their contents, and their dependencies~\cite{MPKB10}. That is, we abstract from the way of creating the results by the use of particular methods and modelling notations and specify \emph{what} has to be done rather than dictating \emph{how} to do something. 

In our experience, the focus on RE artefacts strongly supports achieving the goals of a flexible process that still leads to detailed and, to some extent, (semantically) accurate RE specifications~\cite{MPKB10, MLPW11}. Our process-agnostic focus on what should be created in a project in contrast to how to do something allows us to abstract from the variability in the processes, because the actual creation of artefacts by the use of particular methods in a particular sequence is reduced to the created artefacts, their contents, and their dependencies, all defined in the artefact-based reference model of a company~\cite{MPKB10,MLPW11}. 

\paragraph{Problem Statement.}
Although we have made first steps into the direction of gathering a common understanding about artefact orientation~\cite{MPKB10, MWb13, MLPW11}, the paradigm is still young and it comes too often with various interpretations and manifestations in practice. In fact, little is yet known about how to establish an artefact-based RE approach in practice, which basic concepts have to be taken into account during this establishment, and what benefits as well as shortcomings the paradigm brings. 
This is, however, crucial to steer further evidence-based research within the various research communities and to increase the awareness of the basic principles for the practical application of the paradigm.

\paragraph{Objectives.}

In this article, we aim at providing a consolidated and empirically evaluated view on artefact-based requirements engineering. 

\paragraph{Contribution.}
To provide a consolidated view on artefact-based requirements engineering, we contribute a domain-independent, artefact-based RE approach (the AMDiRE approach), which emerged from six years of experiences in fundamental and evidence-based research. Our contributions are intended to serve more than one purpose:

\begin{enumerate}
\item We introduce the basic concepts of artefact orientation in RE that we have established in fundamental research devoted to this area to lay a terminological and conceptual foundation.
\item We introduce our 6 years of research projects and resulting artefact-based RE approaches to support a common understanding of the various concepts and different interpretations of artefact orientation disseminated into academia and practice.
\item We contribute an artefact-based approach to RE, which uses a tailorable artefact model for domain-independent RE (AMDiRE) as its backbone, and which consolidates our previously developed and evaluated approaches.
\item We share our experiences in the development of artefact-based RE approaches, lessons learnt, and conducted empirical evaluations in industrial contexts. The evaluations also show the practical implications that the different interpretations of artefact orientation have in practice. 
\end{enumerate} 

With our contributions, we aim at supporting researchers as well as practitioners: Researchers  can directly build their fundamental, educational, and evidence-based work upon our artefact model and our experiences to steer their research in a problem-driven manner. Practitioners can directly apply our model in their own socio-economic contexts with the awareness of the benefits and shortcomings of the incorporated concepts.

\paragraph{Delimitations.} 

Instead of preaching the use of one paradigm while neglecting potential benefits of the other, it is our intention to clarify the notion of artefact orientation in RE, draw an outline of its practical application, and discuss the lessons we learnt in recent years. We therefore also discuss the evolution of the paradigm from our experience, and contribute a consolidated approach as a result of various industrial research cooperations. We do not intent to propagate the dogmatic application of artefact orientation for all domains, nor do we claim its valid advantages to hold for all purposes. In fact, we agree with Tell and Babar~\cite{TB12} that, on the long run, we can make use of the benefits of both paradigms while limiting their shortcomings. 

\paragraph{Research Method.} 
\label{sec:researchmethod}

\begin{figure}[hbt]
\begin{center}  
\includegraphics[width=1\textwidth]{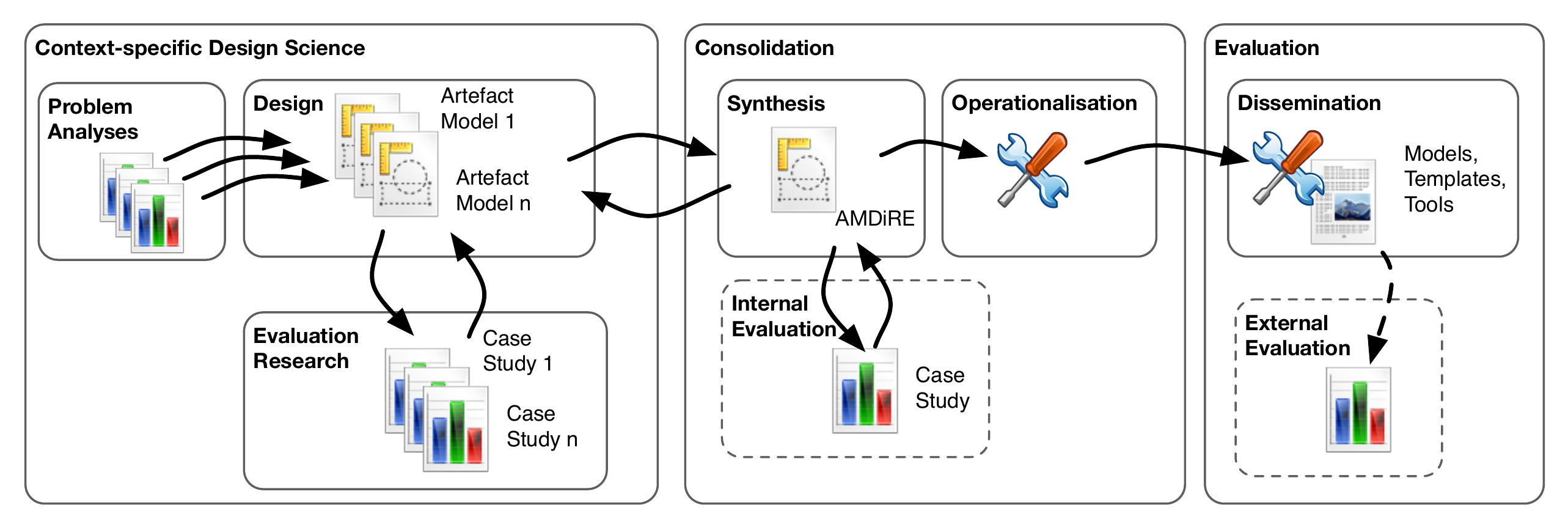}\\
\caption{The applied research method.}
  \label{fig.researchmethod}
\end{center}
\end{figure}

Our contribution at hand, in particular the AMDiRE approach, emerges from a series of different artefact-based RE reference models developed in different research cooperations. For each development in a specific socio-economic context, we followed the principles of empirical design science~\cite{Wieringa09,Wier10a}, i.e. we applied scientific methods in practical contexts to establish an artefact-based RE approach in response to company-specific problems and goals (see also Sect.~\ref{sec:ExperiencesConstruction} where we discuss our general experiences in the construction of those models as well as the procedure we followed). In each project, we  started with a problem analysis (see, e.g.,~\cite{MWLBC10}) to infer a set of improvement goals, before developing particular artefact models via technical action research workshops~\cite{MA12} with our partners from industry. We finally conducted case study research to evaluate each of the developed artefact models w.r.t. the previously determined improvement goals and investigated to what extent we solved the discovered problems. This allowed us to get a deeper understanding on the various characteristics artefact-based approaches can have in dependency to various goals, and what implications those characteristics have when applying the models in practical environments. In Fig.~\ref{fig.researchmethod}, we depict the procedure on the left side.

In a second step, we consolidated the results of the various development and evaluation procedures and synthesised the key concepts of the created artefact models into the AMDiRE approach, which forms the main contribution of this article (see the middle side of Fig.~\ref{fig.researchmethod}). 
So far, we see the resulting AMDiRE approach to be successful, because up to now the concepts from which we inferred AMDiRE have resulted in successful (evaluated) RE reference models leading, e.g.\ to new company-specific RE standards (see, e.g.\ the BISA approach in Table~\ref{tab:CIApproachesOverview}, Sect.~\ref{sec:SummaryCharacteristics}). 

Some of the former models and their empirical evaluations have been published earlier~\cite{MPKB10, MLPW11}. This article presents the consolidation of our work and the actual resulting AMDiRE model on the basis of its development over time.

In contrast to the empirical evaluations of our previous artefact models where we conducted comparative case studies to evaluate to what extent the developments lead to an improvement of previously used activity-based RE approaches, we do not provide such evaluation for AMDiRE in this paper. The reason is that we reached the point where it yet has to be shown whether our approach can be used by others if we are not involved at all, thus, we need an external evaluation independently carried out by unbiased researchers and practitioners not involved in the development of AMDiRE. For this reason, we make our contribution and its operationalisation (e.g. relating models, tools, and evaluation templates) openly accessible~\cite{onlineAMDiRE} and disseminate our results from 6 years of research with the article at hand. This lays the foundation for the final external evaluation, depicted on the right side of Fig.~\ref{fig.researchmethod}.

\paragraph{Outline.}
The remainder of the article is as follows. In Sect.~\ref{sec:FundamentalsAndRelatedWork}, we discuss the work directly related to our contributions, and the gaps we intend to close. We conclude with a discussion of the fundamentals in artefact orientation and the terminology we use in context of this article. In Sect.~\ref{sec:History}, we then introduce the background of artefact orientation resulting from our fundamental, conceptual, and empirical work in this area, i.e. introduce the previously developed artefact models and give a first introduction into the different case studies we conducted with those models. After discussing the synthesis of the models in Sect.~\ref{sec:Synthesis},  we present the AMDiRE approach in detail in Sect.~\ref{sec:AMDIRE}. In Sect.~\ref{sec:ExperiencesEvaluationsLessonsLearnt}, we finally discuss our experiences, our evaluations, and the lessons learnt, and conclude with Sect.~\ref{sec:Conclusion}.

\section{Fundamentals and Related Work}
\label{sec:FundamentalsAndRelatedWork}

We first discuss the areas of activity orientation and then the fundamentals in artefact orientation as well as the terminology used in context of this article. 

\subsection{Form Activity Orientation to Artefact Orientation}
\label{sec:Related Work}

\emph{Activity orientation} is based on the idea of providing an RE reference model as an ordered set of activities and methods, each defining procedures and techniques for a particular purpose~\cite{NE00}, from which project participants can select the appropriate one to design their project-specific RE process. Each activity, e.g. how to apply use cases~\cite{Doerr03}, is performed by a particular role that creates the corresponding artefact type, e.g.\ the requirements specification. Each of those techniques is then placed into a particular sequence of application and used to specify the RE results~\cite{BWHW05}. 

At the organisational level, these activity oriented RE reference models are integrated into activity-based software process models that, for example, rely on the \emph{Software \& Systems Process Engineering Meta-Model} (SPEM)~\cite{SPEM}, such as the  \emph{Rational Unified Process} (RUP)~\cite{KK03}. Approaches that provide means to systematically select and combine methods at project level are addressed, in turn, by the research area of \emph{Situational Method Engineering}~\cite{Brink96, AHV97}. This area can be complemented by \emph{(content-centric) Decision Support Systems}~\cite{RPAW+01}, which contribute approaches to select, classify, and rate a set of alternatives in the choice of methods (and description techniques) according to project parameters. 

Although the importance of a well-defined artefact model is recognised in the area of activity orientation~\cite{FBB08}, the definition of artefacts, their contents, and especially their dependencies is not in scope of available approaches. Braun et al.~\cite{BWHW05} discovered that only 50\% of the analysed approaches include an artefact description at all, while the other 50\% reduce the artefacts to an outcome of self-contained and interconnected methods that produce the artefacts. A first contribution that addresses the incorporation of artefacts into those activity-centric software processes is made by Silva and Oliveira~\cite{Silva} who propose a concept of meta-modeling to define an artefact layer and a process layer for a better organisation of software artefact authoring. They illustrate their approach with a use case specification outline, but do not yet provide a complete artefact model or reference implementation that would provide insights into strengths and weaknesses~\cite{KMS13}. 

Considering the absence of strong empirical work in the area of activity orientation~\cite{PPLB07} and, thus, following a purely argumentative line of reasoning, activity-oriented approaches still have difficulties to overcome the problem of providing a means to support a flexible RE process that guides the creation of consistent RE artefacts. In contrast, when following the principles of \emph{artefact orientation}, we are supposed to define an RE reference model by defining the artefacts, their contents, and their dependencies rather than dictating the way of creating the artefacts, thus, supporting flexibility in the process and the creation of detailed, consistent RE artefacts. First evidence for the benefits of artefact orientation is provided by industrial case studies that evaluate both paradigms in a comparative manner, e.g.~\cite{MLPW11} (see also Sect.~\ref{sec:History}). 

The basic idea of artefact orientation is, however, not new. First artefact models have been proposed as part of checklists and templates for RE, for example, with the VOLERE requirements specification templates~\cite{RR07} or the IEEE recommended practice for software requirements specifications (IEEE std. 830-1998)~\cite{IEEE830}. Those templates provided a first, common understanding on the general contents to be considered in RE artefacts in the form of generic tables of content, but they did not consider the dependencies within and between the contents. The latter is, however, important to support syntactically consistent result structures. 

First content-related dependencies resulting from refinement and decomposition in the modelling concepts are provided by Berenbach et al.~\cite[chp.~2]{BPKR09}. These cover the basic concepts previously developed in a research co-operation between Siemens Corporate Research and Technische Universit\"at M\"unchen (TUM)~\cite{TUM-I0618} (see also Sect.~\ref{sec:History}). They provide an RE artefact model and name the key components for measurable RE artefacts, include a first process guideline, and suggest practices for their elaboration. 

This and similar artefact models enable an understanding about how to structure RE artefacts and how the contents relate to each other. However, those models are limited to general content descriptions rather than providing clear definitions of the modelling concepts used, for example, to create use case models. Thus, they still do not support syntactically consistent result structures. 

This non-exhaustive list of artefact-based approaches already shows that we, as a research community, have developed different views on artefact models depending on their intended purpose. More structure-oriented artefact models, like the one provided by Berenbach et al.~\cite[chp.~2]{BPKR09}, allow for a clear process integration, since a simplified view on the contents of the artefacts can be integrated with process elements like milestones. More content-oriented artefact models, like the one provided by Schaetz et al.~\cite{SFGP05}, focus on (tool-supported) seamless modelling, although a process integration becomes difficult due to the increased complexity in the models~\cite{Schaetz08}.

A meta model for our proposed paradigm is provided in~\cite{MPKB10}. Over the years, we have instantiated this meta model for different domains of applications where the resulting artefact models have been evaluated and disseminated to practice. A discussion of those models is provided in Sect.~\ref{sec:History}. The models had all different contents, but they all relied on the same notion of artefact orientation that we introduce in the following.

\subsection{Fundamentals and Terminology used in Artefact Orientation}
\label{sec:Fundamentals}
In the following, we introduce the basic concepts and the terminology used for artefact-based RE as it results from our previous work~\cite{MPKB10, MLPW11} and as it shall be used in context of this article. The most important terms are listed in Table~\ref{tab:terminology}.

\begin{table}[htb]
\caption{Terminology used in this article.
\label{tab:terminology}}
\begin{center}
\begin{tabular}{p{0.2\linewidth}p{0.7\linewidth}}
\hline
\textbf{Term} & \textbf{Description}  \\
\hline
Project & Software development effort aimed at the construction of a (software) system through the application (execution) of a development process model (see also~\cite{PS06}).\\\hline
Requirements Engineering reference model & Standardised organisational blueprint that includes the description of the generic process (definition) to follow, the artefacts to be generated, as well as roles involved (see also~\cite{PS06, ISO24744}).\\\hline
Process & A process is a series of actions that produce something or that lead to a particular result~\cite{MW2014}. \\\hline
Artefact & Deliverable of major interest that abstracts from contents of a specification document. It is used as input, output, or as an intermediate result of a process step (see also~\cite{MPKB10}). \\\hline
Artefact model & Model that defines a family of artefacts and their dependencies.\\\hline
Method & An information systems development method is likely to include a series of phases with subphases, each having expected outputs (or artefacts); a series of techniques; a series of tools; a training scheme and some underlying philosophy~\cite[p.~44]{AF06}.  \\

\hline \hline
\end{tabular}
\end{center}
\end{table}

Each artefact captures two views: A \emph{structure} view and a \emph{content} view. The structure view captures for each artefact type (e.g., requirements specification) the content items to be considered (e.g., use case model). For each content item, we define the content view via the modelling concepts, e.g., the elements and (content) relations of a use case model and different description techniques that can be used to instantiate these concepts and form the \emph{representation} of an artefact. The structure model is used to couple the contents to the elements necessary to define a process, i.e., to roles, methods, and milestones. 
Regarding the methods and description techniques for creating the contents (e.g.\ UML or natural text), we leave open which one to choose, as long as the contents and relationships proposed by the artefact model are specified.
\begin{figure}[hbtp]
\begin{center}  
\includegraphics[width=1\textwidth]{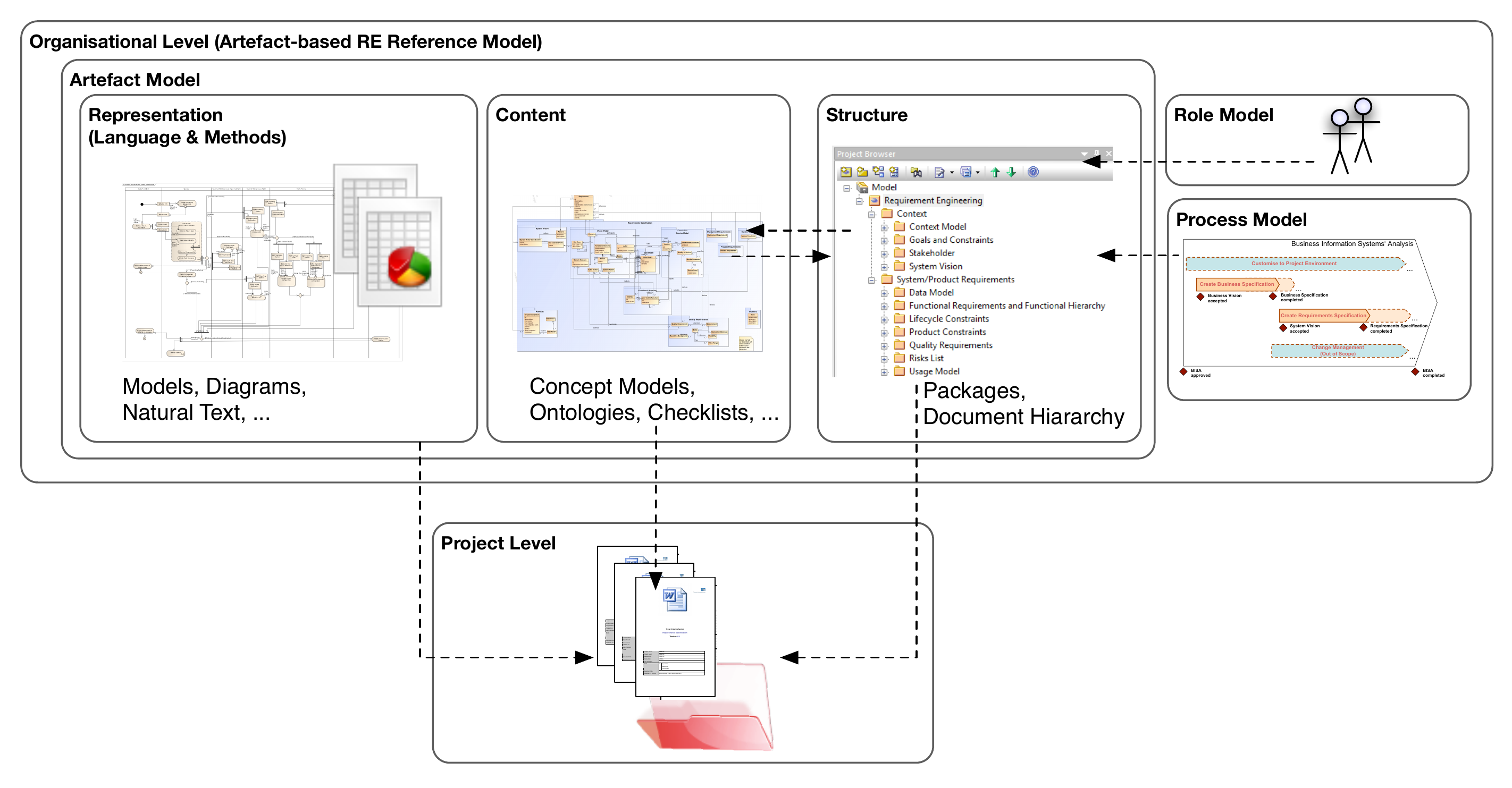}\\
 \caption{Principles of artefact orientation.}
\label{fig.artefact-orientation-principles}
\end{center}
\end{figure}

Same as for activity-oriented approaches, we consider a guiding backbone necessary for artefact-based approaches, which is constituted by the artefact models (see Fig.~\ref{fig.artefact-orientation-principles}, left side). However, instead of defining the artefact-based requirements engineering approach on the basis of interconnected phases, activities and methods, we define the approach on the basis of the artefacts and their dependencies. We define roles and responsibilities for the artefacts to be created as well as the milestones, which define until when to complete, quality assure, and deliver an artefact. This reference model at the organisational level thereby allows to flexibly guide RE at the project level as the way of creating the artefacts is left open to the project participants.

\section{Our Artefact-based RE Approaches and their Synthesis}
\label{sec:History}
\begin{figure}[htbp]
\begin{center}  
\includegraphics[width=1\textwidth]{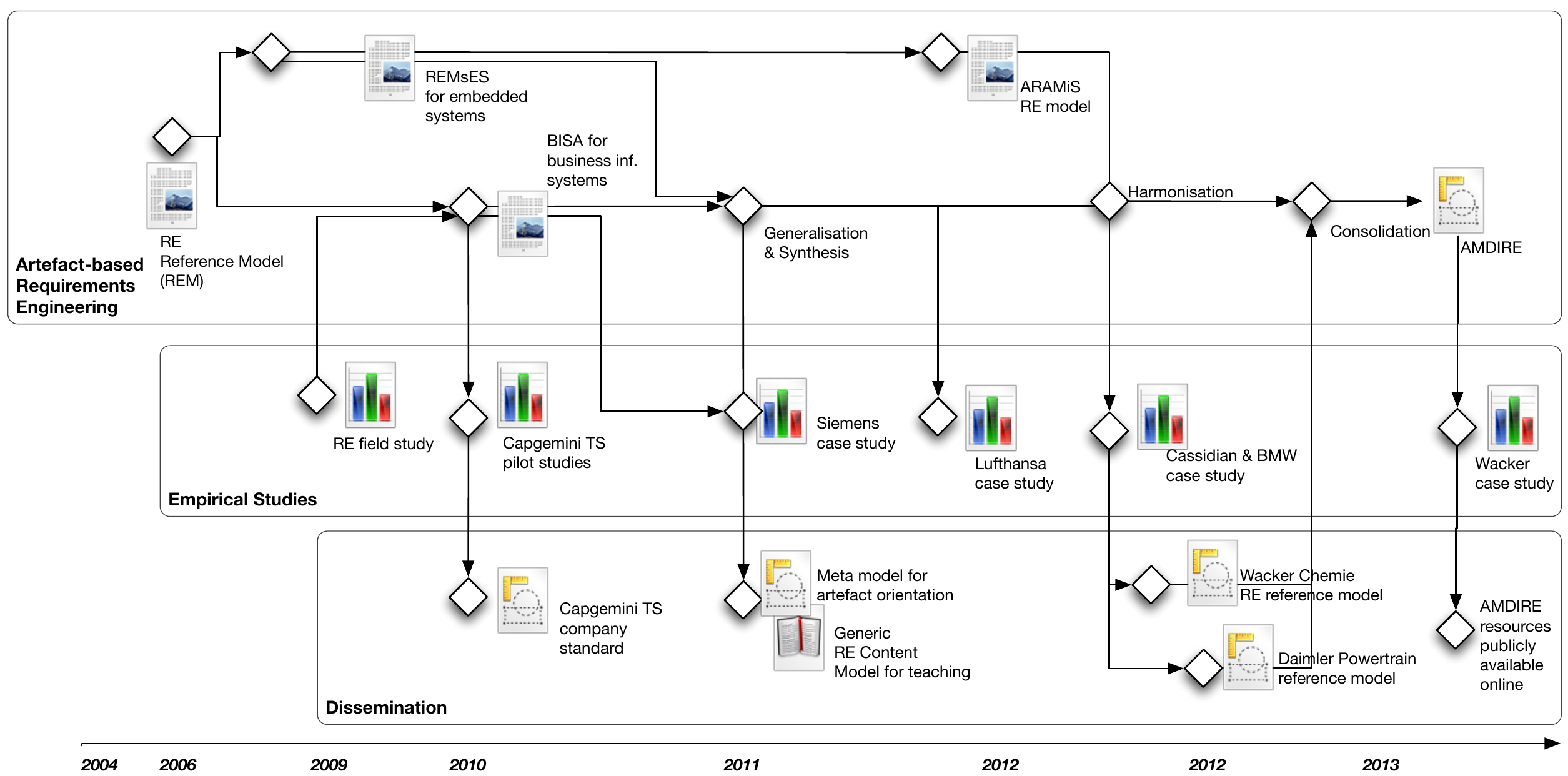}\\
  \caption{Background: Development of artefact-based RE approaches.}
  \label{fig.artefact_history}
\end{center}
\end{figure}

The backgrounds of AMDiRE are various fundamental and conceptual approaches from our previous research. After evaluating and disseminating those approaches into practice during the past 6 years, we synthesise those experiences in the AMDiRE approach. Figure~\ref{fig.artefact_history} illustrates an overview of our previously developed approaches.

The figure is organised into 3 layers. The upper layer shows the developed approaches to RE, followed by the second layer that illustrates major empirical evaluations of those approaches. The positive and negative results gathered from those evaluations served to steer the subsequent development. Finally, the third layer illustrates the dissemination of results (and intermediate results) into academia and practice. 

In the following, we briefly introduce the development, evaluation, and dissemination of our artefact-based RE approaches, before summarising those that serve as a basis for the AMDiRE approach.

\subsection{Overview of Development of Artefact-based RE Approaches at TUM}
Before devoting our research to RE, we, as a research group, investigated the paradigm of artefact orientation in the area of software process models starting from 2004, as depicted in Fig.~\ref{fig.artefact_history}. 
In 2006, we first transferred the basic concepts of artefact orientation to RE. This effort resulted in our first reference model for artefact-based requirements engineering: the requirements engineering reference model (REM)~\cite{TUM-I0618, SFGP05}. REM resulted from a research co-operation between the Technische Universit\"at M\"unchen (TUM) and Siemens Corporate Research. The model defines the structure of goals, requirements and specifications within a proposed taxonomy-based guideline and informally describes dependencies between the elements of the guideline based on proposed refinement principles. Although REM was not intended to capture details of particular application domains, the approach provided a first consolidated view on the previously existing guidelines and checklists available to RE, such as the VOLERE requirements specification templates~\cite{RR07}, the IEEE recommended practice for software requirements specifications (IEEE std. 830-1998)~\cite{IEEE830}, or more practical guidelines such as the one of Wiegers~\cite{Wieg03}. 

A first domain-specific artefact-based approach was developed under the REMsES project~\cite{Braun2010}\footnote{REMsES guide available at \url{http://www.remses.org}}, a research collaboration with partners from academia and industry including BOSCH and Daimler. This project resulted in an artefact model for RE in the automotive domain with a strong focus on contents necessary to specify embedded reactive systems~\cite{PSP09}. The reference model is based on two key concepts: support for abstraction levels and coverage of three content categories. The structure supports requirements engineers in determining which type of model they should use and what kind of abstractions they should consider in a particular project.

In parallel to this development, we worked on another artefact-based RE approach for business information systems analysis (BISA)~\cite{TUM-I0929, Mendez11} as part of a bilateral research co-operation between the TUM and Capgemini Technology Services, the German branch of the Capgemini group for custom software development. The resulting BISA approach is a model-based RE approach that consists of (1) an artefact abstraction model with horizontal abstraction and modelling views, (2) a concept model that defines the possible notions for producing the models, and finally (3) a method description that defines the activities and tasks of the RE process. After two years of development and evaluation in 16 pilot projects the approach became the company standard for RE.

The evaluations of both approaches showed benefits as well as shortcomings. In contrast to the REMsES approach, the BISA approach proved to better support the specification of detailed results due to the detailed concept model, but needed training and coaching. Also, the method descriptions in BISA increased the complexity unnecessarily. 

The subsequent consolidation thus included three steps. First, we integrated the artefact-based approaches into our previously developed software process models~\cite{TUM-I0929} to address a broader audience and to steer further evaluations. Second, we generalised and synthesised our approaches to establish a meta model for artefact orientation. Third, we conducted additional case studies where we applied the consolidated approach in different socio-economic contexts to test the external validity. 

The meta model for artefact orientation~\cite{MPKB10} unifies the different views we had so far on artefact models, including a coarse-grained view on the structure of artefact models as given in development process models and a detailed content view as given in model-based development where we detailed the topics with concrete concept models (successfully introduced by BISA). The coarse view aimed at supporting a flexible process definition, while the concept model aimed at offering guidance for the creation of detailed results. Both views result in our understanding on the constructs necessary to define an artefact, as discussed in Sect.~\ref{sec:Fundamentals}. 

The first case study of the consolidated approach was performed with a street traffic management business unit from Siemens~\cite{MLPW11}. In this case study, we empirically analysed the different benefits and shortcomings of our artefact-based RE approach, but remained aware that the empirical evidence was limited to the particular, sensitive context of our study. The case study, however, was the first one to evaluate the available paradigms to construct RE reference models in a comparative manner by directly comparing our approach with the previously used activity-based one, followed by another study at the Deutsche Lufthansa  (DLH). 

The results in those studies showed us that the views captured in the meta model artefact orientation were valuable to, on the one hand, define a flexible process on the basis of a coarse structure model, and, on the other hand, to guide the creation of detailed results due to the detailed content model. However, the results also indicated that the complexity in the content model implies a higher learning curve\footnote{We understand learning curve in the sense of being related to the power law of practice, such that continued application will lead to learning, as defined and described in~\cite[p.~3/4]{ritter01}.} in the application of the approach in contrast to applying activity-based approaches. 

The subsequently conducted project Automotive, Railway and Avionics in Multi-core Systems (ARAMiS)\footnote{\url{http://www.projekt- aramis.de/}}, a German publicly funded research project where 40 partners from academia and industry worked on an integrated approach for developing cyber physical systems scenarios, resulted in an artefact-based RE approach~\cite{PE2012} with a less complex content model. Subsequent case studies at BMW and Cassidian~\cite{PEM13} followed the same study design as defined for the study at Siemens and strengthened our confidence on the general benefits of artefact orientation, but also that the complexity in the given content model, although necessary to support a high level of detail in the results, hampers its easy applicability. The investigation of these phenomena is in scope of current investigations as part of a family of studies~\cite{PME13}. 

Finally, after preliminary evaluation on an Automatic Cashier System~\cite{Besner13}, the first empirical study on AMDiRE has been completed at Wacker Chemie (reported in Sec.~\ref{sec:amdirecasestudy}), see right side of Fig.~\ref{fig.artefact_history}. Another study investigated the applicability for constructing a RE model in the context of agile methods~\cite{Wiesi13}.
Further studies to contribute to the family of studies~\cite{PME13} are currently in progress. For dissemination, a set of resources --- cheat sheet, Magic Draw plugin, example specifications, and evaluation template --- is available online~\cite{onlineAMDiRE}.

\subsection{Summary of Approaches and Their Characteristics}
\label{sec:SummaryCharacteristics}

Table~\ref{tab:CIApproachesOverview} summarises those artefact-based RE approaches, which serve as a basis for the AMDiRE approach. We take into account their structuring into basic components as well as their contents.

\begin{table}[htbp]
\caption{List of approaches with their evaluations and characteristics.
\label{tab:CIApproachesOverview}}
\begin{center}
\begin{tabular}{p{0.1\linewidth}p{0.15\linewidth}p{0.25\linewidth}p{0.12\linewidth}p{0.18\linewidth}}
\hline
\textbf{Approach} & \textbf{Components} & \textbf{Characteristic} & \textbf{Evaluation} & \textbf{References} \\ \hline 
REM & Artefact model & Structure model & N/A & Model~\cite{TUM-I0618}, \newline Tool~\cite{SFGP05} \\\hline
REMSeS & Artefact model and modelling techniques & Checklists and modelling techniques for embedded systems & Daimler, BOSCH & Model\&Eval.~\cite{PSP09}, \newline \url{http://www.remses.org} \\\hline
ARAMiS & Generic content model & Domain-independent structure model for cyber-physical systems and partial concept model (for tooling) & BMW, Cassidian & Model\&Eval.~\cite{PE2012}, \newline \url{http://www.projekt- aramis.de/} \\\hline
BISA & Artefact model, process elements, customisation approach, & Structure model and concept model for the purpose of process integration & CapGemini (N/A), Siemens & Model~\cite{TUM-I0929, Mendez11},\newline Evaluation (Siemens)~\cite{MLPW11} \\\hline
\hline
\end{tabular}
\end{center}
\end{table}

While the first artefact-based approaches served as initial guidelines, they provided only limited guidance for the content creation as their focus was the establishment of a basic structure model and the inclusion of checklists for the content creation. The BISA approach furthermore incorporated a detailed concept model. This allowed us to support the creation of detailed results as the artefact model made explicit the concepts of an application domain. The structure model additionally supported the process integration, i.e.\ the coupling of the content items to milestones or roles. Other components which turned out to be necessary for application in project environments were a customisation approach as well as tool support relying on the concept model from which we inferred UML profiles. Further information can be found in~\cite{MPKB10, Mendez11}.

\subsection{Synthesis of Established Concepts}
\label{sec:Synthesis}

As discussed in our research method in the introduction (page~\pageref{sec:researchmethod}), we synthesised the established concepts to develop the AMDiRE approach. To this end, we considered the basic components provided by BISA and the content items provided by ARAMiS, which serve as a lessons-learnt-based set of content items relevant for different application domains. 

In order to ensure the applicability of AMDiRE, we made use of the process elements and the customisation approach of BISA that both rely on a structure model. This idea of a plain structure model, in turn, results from ARAMiS, and logically groups modelling concepts constituted by the BISA concept model. 

As AMDiRE is intended to be broadly applicable across application domains, we aggregated, where possible and reasonable, those elements that specify same or similar concepts for different domains into one content item. For instance, AMDiRE includes an element \emph{Domain Model} that includes business process modelling as well as the operational context with hardware and software - the first is relevant to business information systems, the latter for the domain of embedded reactive systems.

In summary, the resulting artefact-based approach shall allow for the specification of detailed results (supported by the detailed concept model) and, at the same time, be easy to use (supported by a simplified structure model). Both views and relating process elements are introduced in the following section.

\section{The AMDiRE Approach}
\label{sec:AMDIRE}

In the following, we describe the AMDiRE approach resulting from the consolidation of the fundamental, conceptual, as well as empirical contributions introduced in the previous section. We first introduce the basic components of the approach and give an overview of the artefact types, the role model and the process model, as well as further constructs used to operationalise AMDiRE in individual socio-economic contexts. We then introduce the artefact model in detail.

\subsection{Artefact Types, Roles, and Milestones of AMDiRE}

\begin{figure}[htbp]
\begin{center}  
\includegraphics[width=1\textwidth]{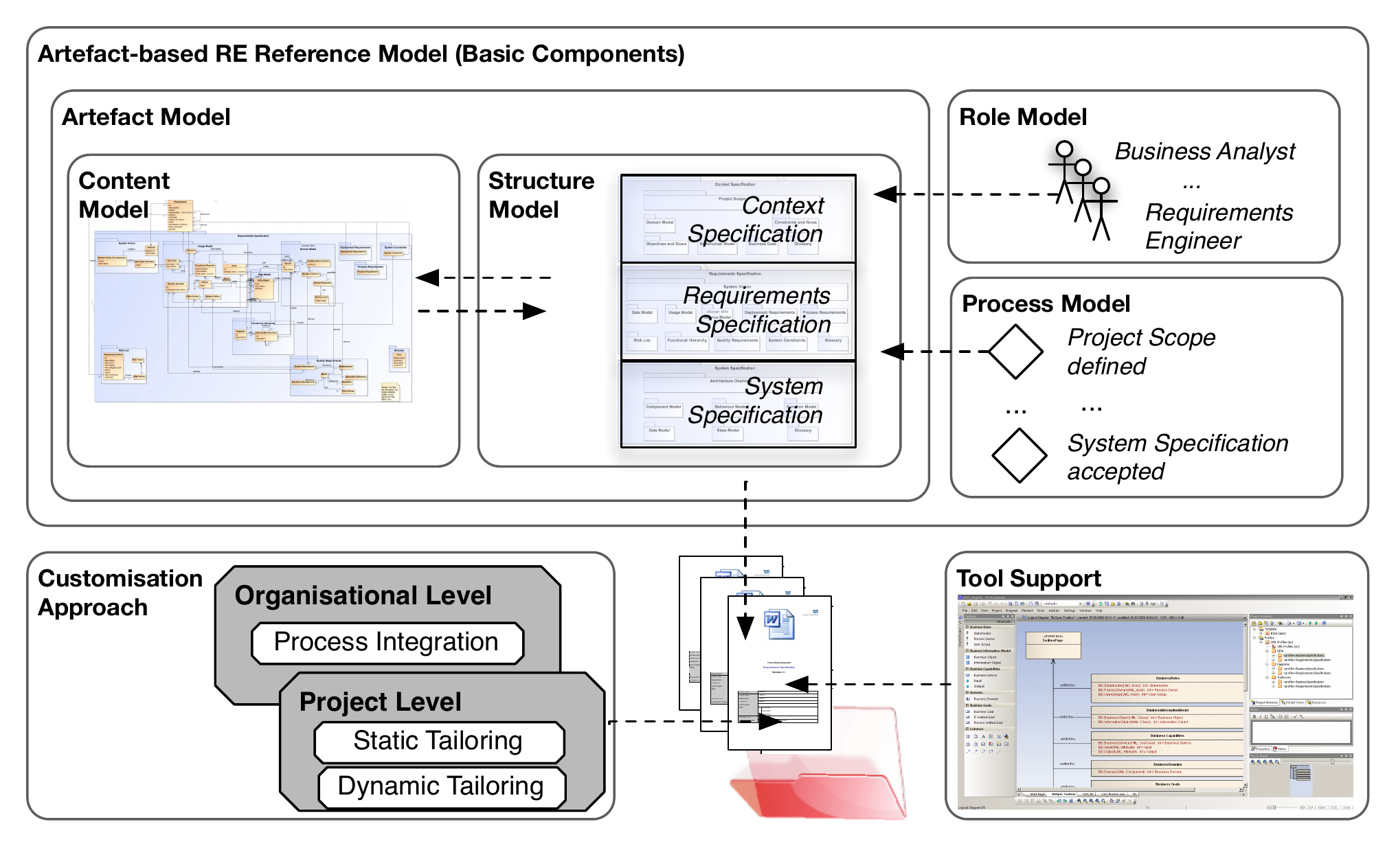}\\
  \caption{Overview of the AMDiRE components.}\label{fig.amdire-approach}
\end{center}
\end{figure}

Figure~\ref{fig.amdire-approach} shows the basic components that build up the AMDiRE artefact-based RE approach. Those components result from our understanding of the artefact-based paradigm as introduced in Sect.~\ref{sec:Fundamentals} and lessons learnt introduced in Sect.~\ref{sec:History}.

The artefact model represents the backbone of the approach and encompasses concepts used to specify the contents of the artefacts. 
This model consists of two basic sub-models: the content model and the structure model. The content model abstracts from the modelling concepts used for a particular family of systems and only scopes the type of information needed. The structure model gives a logical structuring to those concepts and is used for the integration with the role model and the process model.

\begin{figure}[htbp]
\begin{center}  
\includegraphics[width=0.6\textwidth]{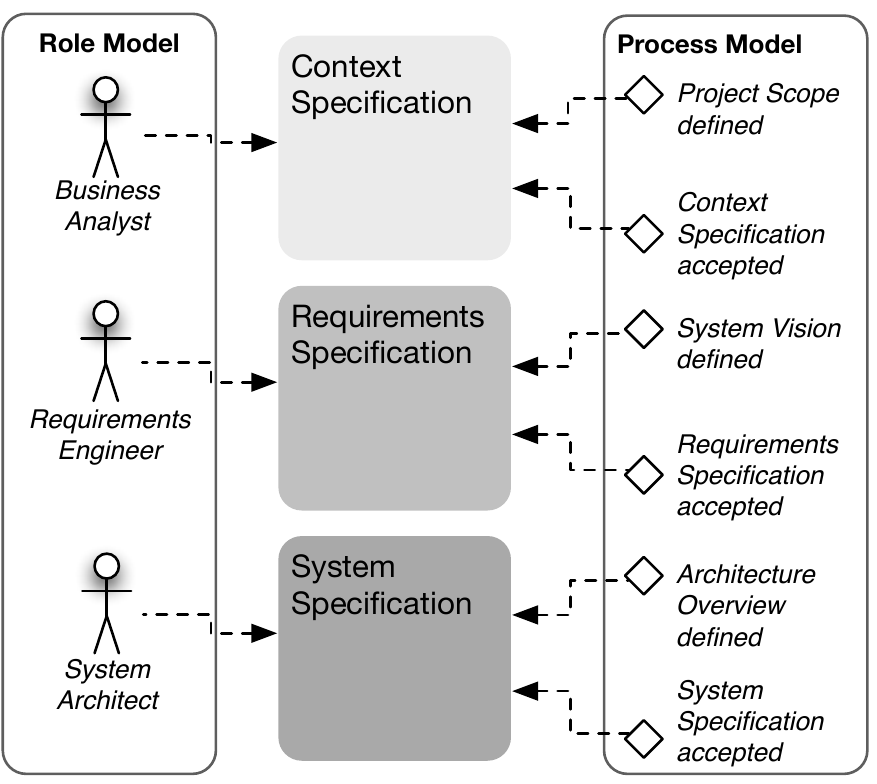}\\
  \caption{Overview of artefacts types, roles, and milestones.}
  \label{fig.amdire_rolesmilestones}
\end{center}
\end{figure}

We distinguish in total three artefact types (Figs.~\ref{fig.amdire-approach} and \ref{fig.amdire_rolesmilestones}):
\begin{enumerate}
\item The \emph{context specification} defines the context of the system under consideration including a specification of the overall project scope, the stakeholders, rules, goals, and constraints as well as a specification of the domain model. The latter comprises, for example, business processes to be supported without, however, defining how the system is intended to be used in context of those processes.
\item The \emph{requirements specification} comprises the requirements on the system under consideration taking a black-box view on the system, i.e.\ we specify requirements from a user's perspective without constraining the internal realisation of the system.
\item The \emph{system specification} finally comprises a glass-box view on the internal realisation of a system including a logical component architecture and a specification of the behaviour realisation with, e.g., functions and interfaces. While we consider the context and the requirements specification to address the problem space, the system specification addresses the solution space and is the interface to tie in with the design phase.
\end{enumerate}

Figure~\ref{fig.amdire_rolesmilestones} shows the artefact types in relation to roles and responsibilities (left side) and in relation to milestones (right side) which we use to integrate the model into a process. 

For each artefact type, we define one particular role, which has the responsibility for an artefact type, independent of other potentially supporting roles provided by the software process model (e.g., quality manager), and independent of whether same persons are assigned to different roles in a project.
\begin{enumerate}
\item The \emph{Business Analyst} has the responsibility for the context specification and is expected to have the necessary domain knowledge, e.g.\ regarding the business processes, typical stakeholders, or constraints and rules.
\item The \emph{Requirements Engineer} has the responsibility for the requirements specification and serves also as a mediator between the business analyst and the system architect.
\item The \emph{System Architect} has the responsibility for the system specification and is expected to have technical knowledge. In dependency to the application domain, we can further distinguish between a role for the logical architecture and a role for the technical architecture (e.g.\ in the area of business information systems). 
\end{enumerate}

For each artefact type, we furthermore define two milestones. The first milestone defines the point in time in which the first content item is defined, thus, reflecting a certain maturity of the content in the artefact as the first content items serve the purpose of a summary for subsequent contents. For instance, the system vision in the requirements specification comprises an overview of the major use cases; its definition and agreement indicate that the use cases are sufficiently defined to be further refined and modelled and, thus, allowing, for example, for first cost estimations based on function points. The second milestone of each artefact indicates the point in time when an artefact is finalised, respectively formally accepted.

Those milestones serve the purpose of a process integration and instantiation as they give us the opportunity to formally embed the artefacts into project-specific decisions. These decisions are to be taken at a specific point in time, such as when to conduct first cost estimations, when changes in the requirements should be formally defined via change requests, or when to take the contents in the specifications for a project classification and customisation (tailoring). 

\subsection{AMDiRE Artefact Model}
In the following, we introduce the refinement principles over the three levels of abstraction by giving an overview of the content-related dependencies between the artefact types. Afterwards, we outline the content model. 

The artefact model is specified using the following notational aspects of UML class diagrams:
\begin{itemize}
\item We denote the hierarchical structuring of the structure model with packages.
\item For the definition of the content model, we use a class diagram.
\item For content items that are crucial for only a specific application domain, but irrelevant for another, we use the stereotype \emph{$<<$Domain$>>$}, such as business process models being crucial for the domain of business information systems, but irrelevant for the domain of embedded reactive systems.
\end{itemize}

\subsubsection{Refinement Principles and Artefact Dependencies}

\begin{figure}[htbp]
\begin{center}  
\includegraphics[width=1\textwidth]{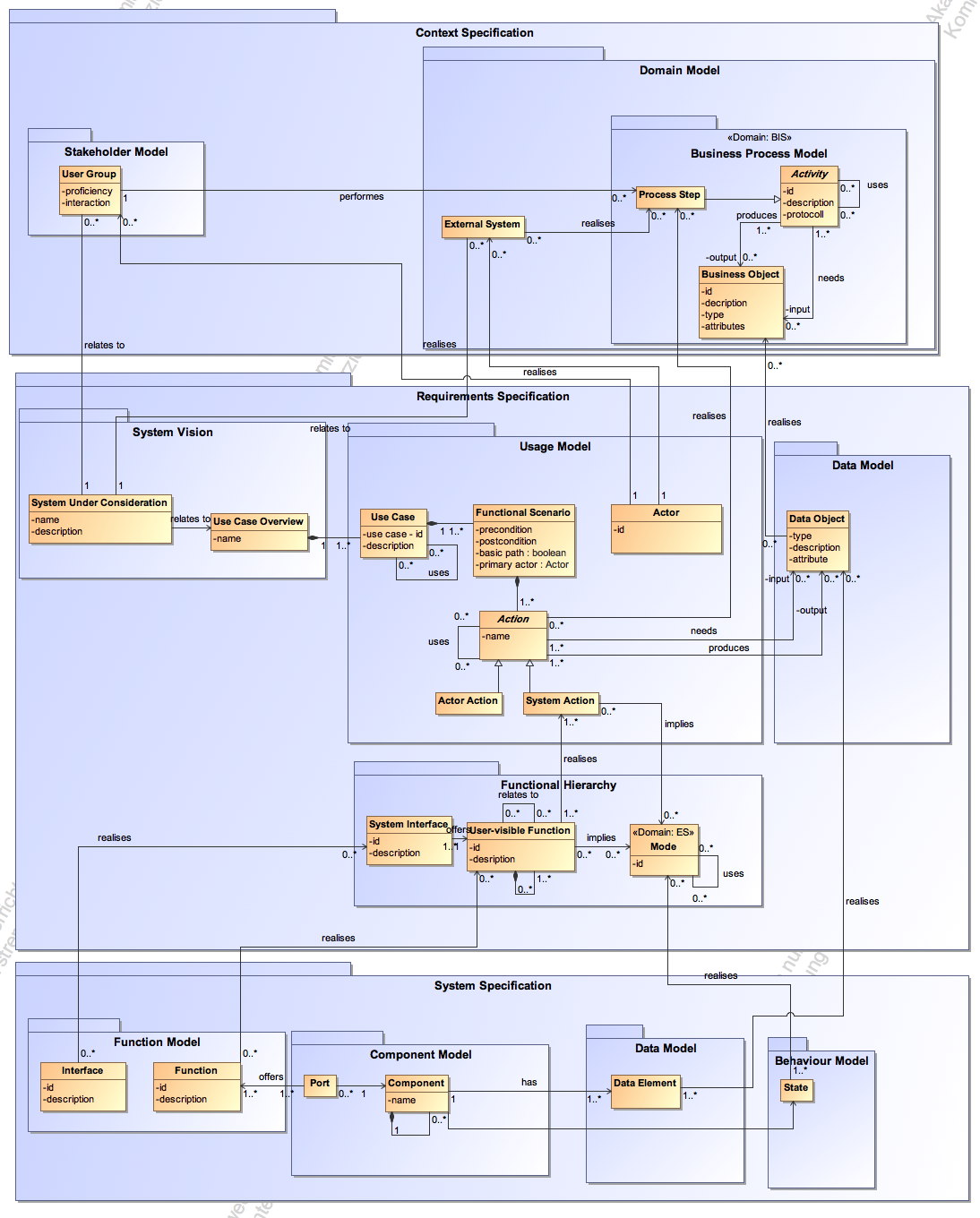}\\
  \caption{Refinement and Realisation Principles in AMDiRE (complete relations are visible in the detailed subfigures in the appendix).}
  \label{fig.amdire_topdown}
\end{center}
\end{figure}

Figure~\ref{fig.amdire_topdown} organises the three artefact types in a top-down hierarchy reflected in the three previously introduced levels of abstraction (see also Fig.~\ref{fig.amdire_rolesmilestones}) and shows the refinement principles we use when modelling requirements and system properties.
For reasons of complexity, we intentionally refrain from a complete overview of the artefact model and instead focus on selected concept types to introduce the content-related dependencies. 
 
In the context specification, we capture behaviour in form of stakeholders performing selected processes. We specify, for example, a business process model that dictates functional behaviour by a set of process steps interrelated in a causal manner. 

In the requirements specification, we select those steps to be supported by a system and specify how the system is intended to be used in interaction with the user groups. The content-related dependencies between both artefact types is given as follows:
\begin{itemize}
\item \emph{Actors} to which we refer in usage models (e.g.\ in use case models specified via UML activity diagrams) realise either \emph{User Groups} or \emph{External Systems}.
\item \emph{Data Objects} to which we refer to specify which information is used as input or output to our system from a black-box perspective realise selected \emph{Business Objects}.
\item \emph{Actions} to which we refer when specifying external system behaviour (e.g.\ via usage models) realise selected \emph{Process Steps} by either defining actions an actor performs or actions a system shall automatise.
\end{itemize}

In the system specification, we then specify how the system will realise the functional external behaviour within a component architecture and its internal behaviour. To this end, we define the following content-related dependencies:

\begin{itemize}
\item  \emph{System Functions}, which are provided by components, realise user-visible functions, i.e.\ those \emph{System Actions} in an \emph{Usage Model} specified in the requirements specification. Same holds for the \emph{System Interfaces}, which realise the identified (typed) \emph{Interfaces}. The realisation dependencies are limited to the external interfaces and functions as we enrich the system specification during design activities with additional internal functions and interfaces between the components of a (logical and/or technical) component architecture.
\item \emph{Data Elements}, which are allocated to specific components and which are processed by system functions over their typed interfaces, realise the \emph{Data Objects} specified in the requirements specification. 
\item \emph{States} realise the \emph{Modes} in a requirements specification and form system behaviour by interrelating the different states of a system via state machines. Similar as it is the case for the transition from the context to the requirements being of interest for business information systems, the transition at hand is of interest when addressing the domain of embedded reactive systems where we identify the relevant states during requirements engineering (reflected in modes). 
\end{itemize}

\paragraph{Refinement of Quality-related Properties} Further elements relevant for the transition between two levels of abstraction (not shown in Fig.~\ref{fig.amdire_topdown}) are those we use to specify quality and quality requirements as part of more general non-functional requirements. In our understanding, non-functional requirements cover system-related quality aspects as well as requirements on properties of the development project specified, e.g., via \emph{Process Requirements}~\cite{Glinz07}. Quality is a multifaceted topic with different views of the term \emph{Quality}~\cite{1984_garvind_product_quality, KP96}, and no commonly accepted definition~\cite{Glinz07}. To avoid ambiguities and to be precise with what we consider as \emph{Quality}, we explicitly refer to a quality definition model. Due to the focus we have in RE on specifying activities with business processes and use cases, we rely on the activity-based quality definition model~\cite{DJLW09, Deis09} by TUM, which is based on early efforts of Boehm et al.~\cite{BBKL+78} and McCall et al.~\cite{CM78}. 


The basic idea in this model is to define quality via a set of system properties and their associations to activities carried out during the use of the system~\cite{Deis09}. For AMDiRE, this means that we define quality via (1) abstract goals~\cite{Cock00} over different levels of abstraction to motivate the refinement of behaviour. This behaviour is specified via (2) generic scenarios that define which non-functional activities the system shall support (e.g., activities the administrator carries out), and, finally, refine those scenarios to (3) assessable quality requirements, which then are used to motivate design decision in a component architecture. 
Consequently, we do not follow a strict separation of concerns regarding quality, but see the notion of quality to affect and define behavioural properties of a system and structural properties as well. 



\subsubsection{AMDiRE Content Model}

The AMDiRE content model is structured into three artefact types that encompass over 70 elements and various relations. For the sake of clarity, we provide a simplified view on the content model while detailed information of all content items and the underlying concept model are provided in Appendix~\ref{sec:ContentModel}. Figure~\ref{fig.amdire_picto} illustrates the content model and shows a sketch for each content item. 
\begin{figure}[htbp]
\begin{center}  
\includegraphics[width=0.7\textwidth]{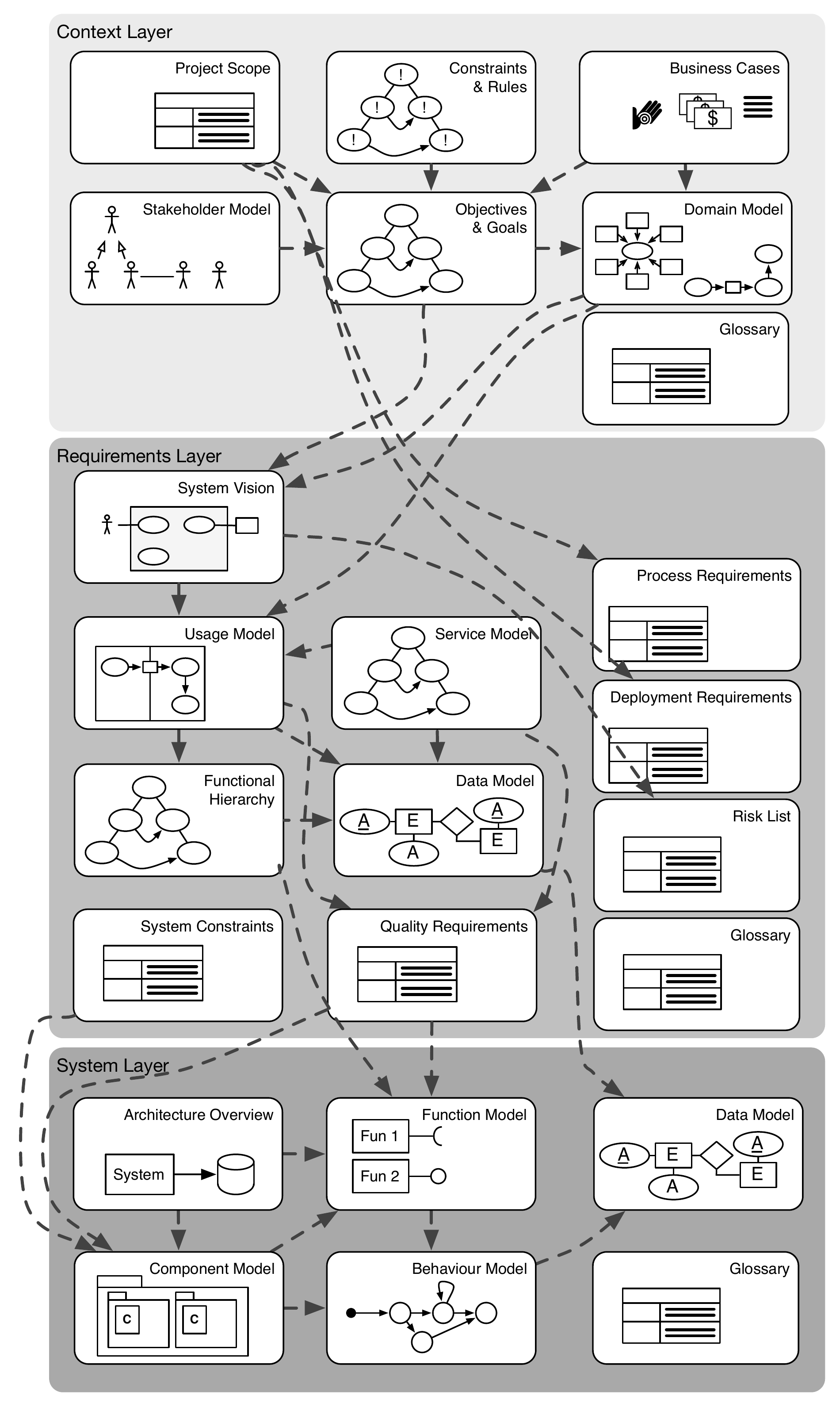}\\
  \caption{Overview of AMDiRE Content Model in a simplified manner.}
  \label{fig.amdire_picto}
\end{center}
\end{figure}
For reasons of reducing complexity, we likewise depict only a subset of the dependencies between the content items (for the complete list, see Appendix~\ref{sec:ContentModel}).
As introduced in the previous sections, AMDiRE relies on a refinement notion for functional as well as non-functional modelling concepts. 

Starting at the top, in the \emph{Context Specification}, the \emph{Project Scope} defines the relevant problem to be addressed by a project and the primary scope. The \emph{Stakeholder Model} is used to capture the most relevant stakeholders and the relationships and are used as a central definition of key reporting lines and one important rationale for requirements and goals. \emph{Goals} are specified, e.g., in a graph form, and serve as a means to steer the specification of a business process model in the domain model. 
The \emph{Domain Model} provides information on the operational environment. 

In the \emph{Requirements Specification}, the \emph{System Vision} defines the basic idea of the system under consideration and the stakeholders of a project agree on a system scope (major features and use cases) as well as its boundaries specified via a context diagram or a rich picture. To capture functional behaviour in the \emph{Usage Model}, we define for each identified use case how future users intend to use the system in interaction.
A \emph{Service Model} is used as a complementary means to define which services the system shall offer -- in contrast to a use case model not necessarily defining the relation to actors but, instead, the causal relations between the services. We further use the system-supported actions in a use case model, e.g., specified via UML activity diagrams, to select candidates for user-visible system functions, which we structure and refine in a \emph{Functional Hierarchy}. This hierarchy builds the point of entry into the system specification. 

In the \emph{System Specification}, we finally allocate the \emph{Functions} of a functional hierarchy to \emph{Components}, define their syntactic interfaces and their internal \emph{Behaviour} with, for example, automata. This behaviour specification also serves the identification of the (typed) entities defined in relation to each other in a \emph{Data Model}.

\subsubsection{Customisation Approach and Tool Support}
\label{sec:CustomisationandTool}
The operationalisation of AMDiRE in a particular socio-economic context is done by two means. In a tailoring approach, AMDiRE is customised over two levels of abstraction:

\begin{enumerate}
\item Organisational level: At the organisational level, we consider the company-specific customisation of AMDiRE, i.e.\ the process integration into an individual organisational reference software process model such as RUP (or a company-specific derivate).
\item Project level: At the project level, we consider the instantiation of AMDiRE (initial artefact creation and decision for specific content items, assignment of roles and definition of milestones), which is known as static tailoring, and the dynamic tailoring during project execution. The latter considers the situation-aware creation of content items in dependency to project-specific situation that affect the need to create particular items and ones that affect the possibility to create particular items. An exemplary set of project influences and their dependency to RE artefacts can be found in~\cite{MWLBC10}.
\end{enumerate}

Both customisation at organisational level and at project level are introduced in~\cite{Mendez11}. In order to further operationalise AMDiRE, we rely on tool support. 
\begin{figure}[htbp]
\begin{center}  
\includegraphics[width=1\textwidth]{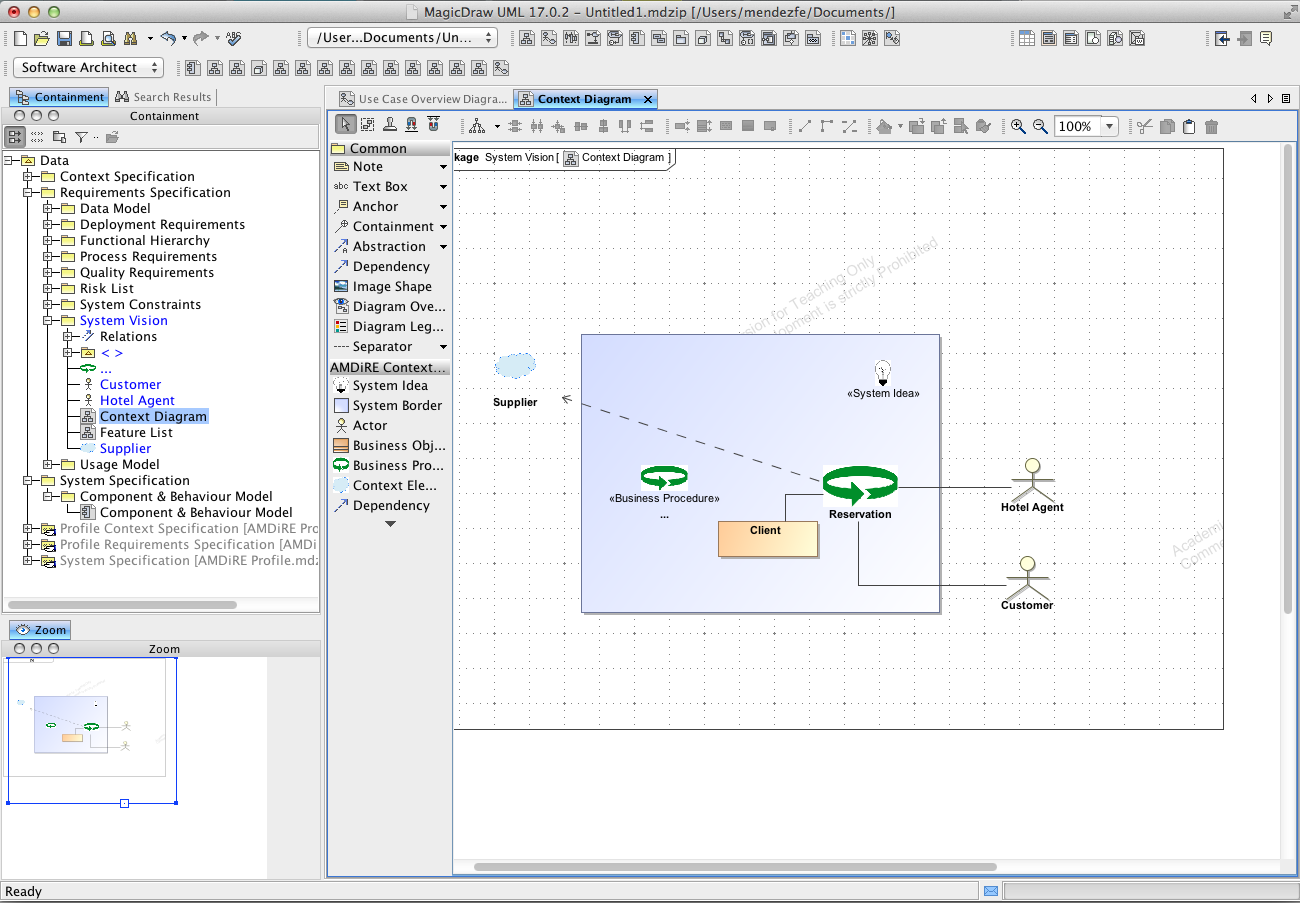}\\
  \caption{Screenshot of model-based tool support at the example of a context diagram.}
  \label{fig.amdire_tool}
\end{center}
\end{figure}
This is currently prototypically realised with an extension of the model-based CASE tool MagicDraw~\footnote{\url{http://www.nomagic.com}} by defining an UML profile based on the content model provided by AMDiRE (see Fig.~\ref{fig.amdire_tool}).
The tool extension is available online~\cite{onlineAMDiRE}.

\subsubsection{Example Application}
This section shows an illustrative excerpt of example models for a fictitious Automatic Teller Machine (ATM). The example was developed for a Requirements Engineering lecture at Technische Universit{\"a}t M{\"u}nchen following the artefact model of AMDiRE and using the tool extension depicted in the previous section. The complete example models can be obtained together with the tool extension from our online resources~~\cite{onlineAMDiRE}.

\begin{figure}[htbp]
\begin{center}
\includegraphics[width=\textwidth]{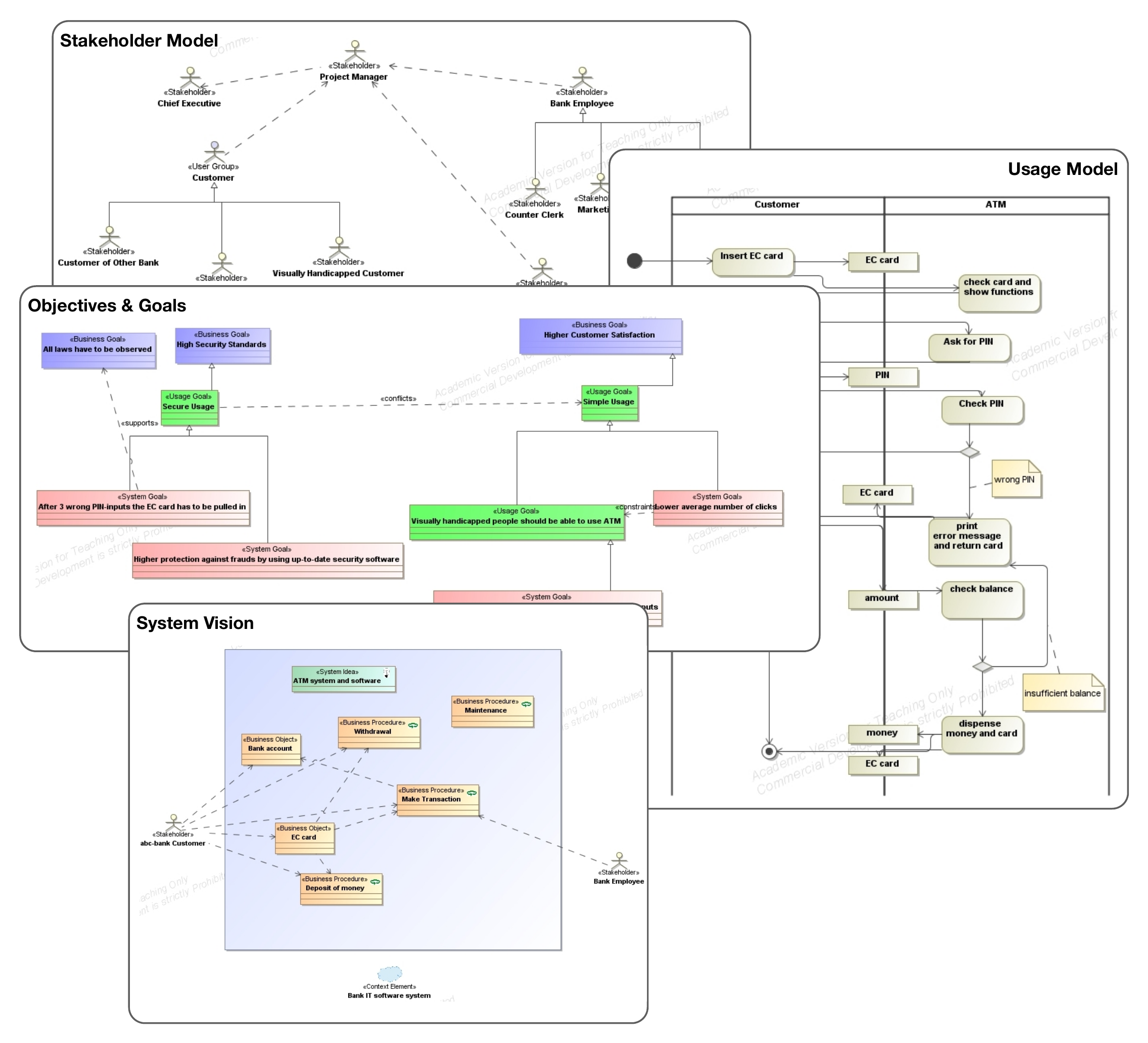}
\caption{Illustrative excerpts of the ATM example}
\label{fig:example}
\end{center}
\end{figure}

Figure~\ref{fig:example} shows the exemplary system vision, the stakeholder model, the goal model, and a scenario from one of the use cases created.
The system vision is the agreed-on vision of the ATM that denotes the system border and the most central features, like \emph{withdrawal} and \emph{transaction}. The stakeholder model includes business stakeholders and user groups in various hierarchies. The goal model includes business goals like \emph{higher customer satisfaction}, usage goals like \emph{visually handicapped should be able to use ATM}, and system goals like \emph{high protection against fraud}. The activity diagram illustrates the scenario of a user \emph{withdrawing money} from the ATM.

\section{Experiences, Evaluations, and Lessons Learnt}
\label{sec:ExperiencesEvaluationsLessonsLearnt}

We have developed a number of artefact models, conducted case studies with different companies, and gained different experiences. In the following, we first discuss the experiences we gained throughout the construction of artefact models. Furthermore, we discuss the industrial evaluations we performed. This shall give a picture of benefits and shortcoming in the construction and the application of the various notions in artefact-based RE. We finally conclude with a discussion of general lessons learnt in the construction and application of artefact orientation in industrially hosted environments.

\subsection{Experiences in the Construction of Artefact-based RE Approaches}
\label{sec:ExperiencesConstruction}
Over the last years, we have performed a series of different research cooperations in the field of artefact-based RE improvement where we developed company-specific artefact-based RE approaches. 

\begin{figure}[htbp]
\begin{center}  
\includegraphics[width=1\textwidth]{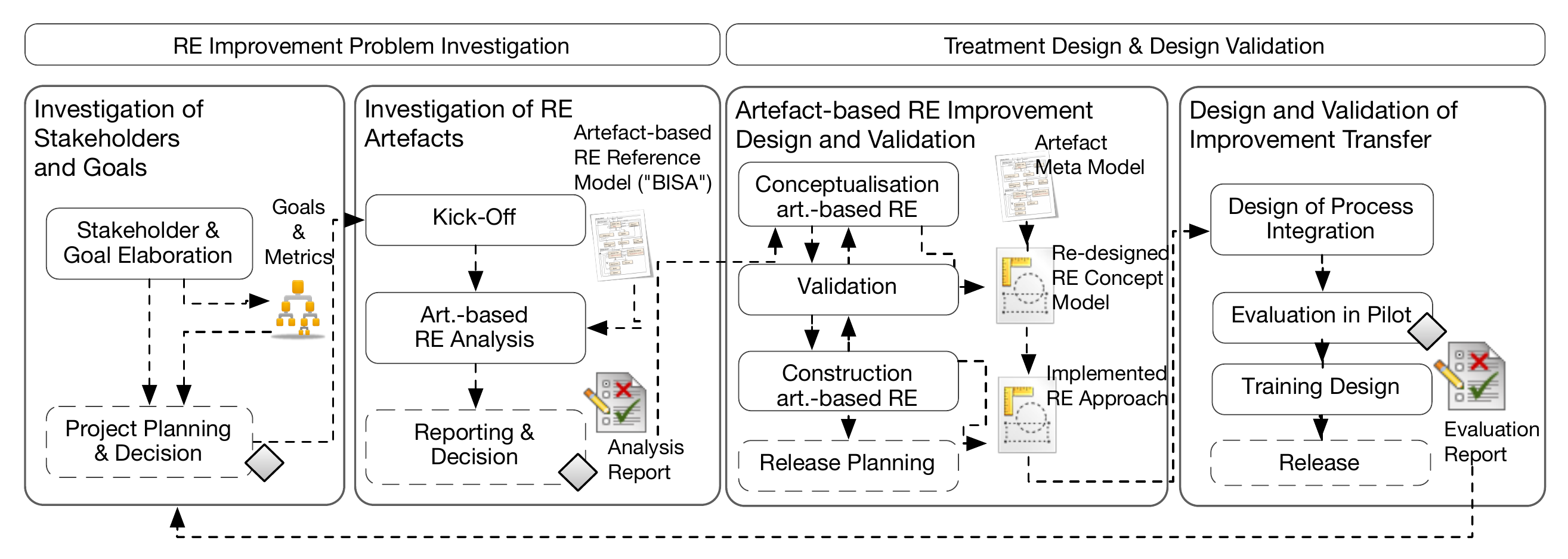}\\
  \caption{Procedure for constructing artefact-based RE approaches.}
  \label{fig.reimprovement}
\end{center}
\end{figure}

The research approach that has proven feasible for us throughout a number of projects is to follow a problem-driven exploration where we apply concepts of empirical design science~\cite{Wieringa09,Hevner04} to build a company-specific RE approach as part of an RE improvement endeavour. Figure~\ref{fig.reimprovement} illustrates this research approach, separating two basic phases: The actual problem investigation, and the design and validation phase. 

Briefly summarised, we begin with a kickoff where the most important stakeholders for the artefact model and their roles are identified, followed by semi-structured interviews with these. Subsequently, we review requirements documents that have been produced by the company according to their current reference model and/or practice and compare these to our content model (gap analysis). The differences found in that comparison are discussed with a representative set of stakeholders in a workshop where we ask for reasons for specific elements we had expected but were not being documented and for extra elements that were documented but not present in our concept model. This allows for a first evaluation where our reference model should deviate for the later elaboration of the to-be model. 

On this basis, we develop an artefact model that is tailored to the needs of the company by performing a series of action research workshops, and we perform a pilot study and/or a comparative case study to evaluate whether the artefact model brings the desired improvement for their requirements engineering practice. A detailed explanation of our approach to construct artefact models is provided in~\cite{MWb13}.

Apart from the methodological experiences we made when building up various artefact models (introduced in~\cite{MWb13}), for example, regarding the use of action research workshops to support knowledge transfer~\cite{MA12}, we gained experience in structuring artefact models at a more syntactic layer. 

In~\cite{MPKB10}, we already discussed first experiences regarding the notion of artefact orientation and inferred a meta model for the paradigm. We discussed that we need to consider different views on artefact models depending on company-specific objectives. Those objectives affect the way artefact models are structured (e.g.\ by defining a detailed concept model via a data model or by defining the content model via a checklist). The models can have different levels of detail according to how much detail the company wants in the model, leading to more complexity, which affects ease of use. Consequently, an increased level of detail requires a higher learning curve by the applying company. This trade-off between higher level of detail and ease of use has to be addressed.

In Table~\ref{tab:CIApproachesOverview} (Sect.~\ref{sec:SummaryCharacteristics}), we selected previously developed artefact-based RE approaches and their characteristics. In Table~\ref{tab:ArtefactCharacteristics}, we structure different approaches in a similar way and summarise the main objectives of the company and their effects on the establishment of the artefact models, as well as the effort spent in their development (in person months). 

\begin{table}[htbp]
\caption{Artefact models, their objective and resulting characteristics.
\label{tab:ArtefactCharacteristics}}
\begin{center}
\begin{tabular}{p{0.02\linewidth}p{0.18\linewidth}p{0.3\linewidth}p{0.38\linewidth}}
\hline
 & \textbf{Approach} & \textbf{Objectives} & \textbf{Characteristics \& Effort} \\ \hline 
\#A & BISA / Quasar Requirements (Capgemini TS) & Process integration and seamless modelling & Structure model and content model defined via a data model and conformance constraints, 112 PM \\\hline
\#B & Generic Content Model (ARAMiS) & Common terminology across the 40 project partners and a concept model for tool support and guidance & Structure model and implemented concept model / UML profile, 20 PM \\\hline
\#C & Artefact Model for Wacker Chemie & Template-oriented checklist to integrate RE artefacts into quality assurance & Structure model and templates, detailed concept model for use cases and the relation to test cases, and process elements, 8 PM \\\hline
\#D & Powertrain Artefact Model (Daimler) & Clarifying the notion of requirements contents & Structure model and detailed concept model without roles, milestones or other process elements, 6 PM \\\hline
\hline
\end{tabular}
\end{center}
\end{table}

Top to bottom in Table~\ref{tab:ArtefactCharacteristics}, we can see
\begin{compactitem}
\item An increase in the details the artefact model incorporates by its underlying content models. Those range from detailed data models in case of the BISA approach to support seamless modelling to artefact models that incorporate checklists and selected concepts to clarify the notion of selected contents (e.g., use cases).
\item A decrease of the effort spent in the construction of those models. Taking into account our own learning curve during the construction of those models, we still have efforts that differ by a factor of 18. Apart from light-weight versus heavy-weight models, this is mainly due to the up-front budgeting and time framing for the projects.
\end{compactitem}

The simpler models developed in context of Daimler (\# D) and Wacker Chemie (\#C) follow a similar objective where we established the artefact-based RE approach to clarify the basic RE concepts and corresponding terminology. Those models serve the purpose of giving a quick overview of elementary modelling concepts such as use case models, their allocation to requirements-specific artefacts and their dependencies to surrounding development phases. At Wacker, the process integration could explain the additional 2 PM. During this integration, we defined roles and responsibilities, a quality assurance process and the integration of a tailoring profile (see Sect.~\ref{sec:CustomisationandTool}). The effects of following those objectives can be observed, however, especially in the way the artefact models were created. In scope were the structure models to define which content items were of general interest. For each content item, we provided a brief guidance in form of a checklist and only specified the concept model in detail where necessary (e.g.\ for the use case model to define the dependency to testing).

In contrast, the objectives followed by the two more detailed models (\#A and \#B) were mainly to support seamless modelling and tool support. The implication for the artefact model was that the content models were specified in full via a data model to define which concepts and relations have to be considered during the artefact creation. Both concept models were also enriched by a structure model to ease the process integration.

The effort in the detailed models mostly arose from:
\begin{compactenum}
\item Terminological and conceptual discussions at workshops, frequent action research workshops to realise the concepts, and a longer review and release process at the partners' sites as the approaches had not anymore the character of being a checklist.
\item The development of training and coaching material as well as tool support.
\item The supervision of pilot projects and coaching during those pilot studies.
\item More general, a missing willingness to organisational change we could observe in the projects as project participants were confronted with complex models and needed a longer learning curve to understand the basic concepts and tailoring mechanisms.
\end{compactenum}

\subsection{Evaluations: Results from Industrial Case Studies}
\label{sec:Evaluations}

We performed a series of evaluations with the previously developed artefact models on which AMDiRE relies. We complement this evaluation with a case study where we directly used AMDiRE as a reference model to establish a company-specific artefact-based RE approach following the principles described in the previous Sect.~\ref{sec:ExperiencesConstruction}.

\subsubsection{Summary of Previously Conducted Comparative Case Studies}

We have performed three comparative, industrial case studies that we briefly discuss with regard to their results. These studies each evaluated an artefact-based approach in direct comparison to the formerly used activity-based approach and were performed with partners at Siemens, BMW, and Cassidian. Table~\ref{tab:CaseStudiesOverview} provides the type of studies and their major results.

In the following, we summarise those previously published case study designs and results relevant to the context of the article while details on the study designs and the full results can be taken from references listed in Table~\ref{tab:CaseStudiesOverview}.

\begin{table}[htbp]
\caption{Comparative industrial case studies.
\label{tab:CaseStudiesOverview}}
\begin{center}
\begin{tabular}{p{0.2\linewidth}p{0.1\linewidth}p{0.08\linewidth}p{0.45\linewidth}}
\hline
\textbf{Study} & \textbf{Approach} &\textbf{Partner} & \textbf{Results} \\ \hline 
Traffic Lights Managment System & \#A & Siemens \newline (2011) & Improvement of, inter alia, syntactic consistency and completeness, ease of use, effectivity, structuredness and ease perception~\cite{MLPW11}\\\hline
Intelligent Infotainment Systems (IIS) & \#B & BMW \newline (2012) & Improvement of effectivity, productivity, and adequacy~\cite{PEM13, PE2012}\\\hline
Unmanned Aircraft System (UAS) & \#B & Cassidian \newline (2012) & Improvement of ease of use, effectivity, unambiguousness and adequacy~\cite{PEM13, PE2012}\\\hline
\hline
\end{tabular}
\end{center}
\end{table}

\subsubsection{Case Study Designs Overview}

As shown in Table~\ref{tab:ArtefactCharacteristics} and in Sect.~\ref{sec:History} introducing the previously developed models, we initiated our case studies with smaller studies at Capgemini TS where we evaluated the approach in 12 internal pilot projects. As those studies remain unpublished due to a non-disclosure agreement, and because we aimed at increasing the external validity, we applied our approach \#A in nother socio-economic contexts. After this evaluation, we continued the development of our artefact models leading to approach \#B which, again, we evaluated to test its sensitivity in a socio-economic context. 

Overall, we conducted three case studies where we evaluated two approaches (see Table~\ref{tab:CaseStudiesOverview}). We evaluated approach \#A at Siemens and approach \#B at Cassidian and at BMW. The objectives of the case studies were defined, at the time of conducting the studies, according to the improvement goals of the projects. That is, in dependency to the defined goals, we formulated the research questions and the evaluation criteria. However, the overall study design remained the same for all studies to allow for a comparability of the results. 

In each study, we conducted a series of workshops with our industry partners and specified the requirements for a selected subsystem following the established artefact-based RE approach. We then conducted an assessment workshop with interviews where the subject evaluated the created RE artefacts as well as the way of working in direct comparison to the artefacts and the process dictated by the (activity-based) RE reference model previously used in the same context. For the assessment, we intentionally conducted one interview session with all involved participants to enable joint discussions among the participants and to get a final agreement on the ratings (without checking the inter-rater consistency). In all case studies, we involved a requirements engineer (in case of Siemens represented by the role of a product manager) and the overall project lead responsible for broader project management activities. 

The rating was conducted by answering a questionnaire with a series of open and closed questions. The closed questions aimed at rating single criterion, such as \emph{ease of use} on a Likert Scale while the open questions served to justify the ratings. The choice of the criteria was performed in discussion with the project participants according to the (improvement) goals and can be taken from the Kiviat diagrams in subsequent result sections.

\subsubsection{Case Study Results Overview}
In the following, we summarise the results from the previously conducted case studies relevant to the context of this article. 

\begin{figure}[htbp]
  \centering
  \subfigure[Internal rating]{
	\label{fig:siemens1}
	\includegraphics[width=0.8\columnwidth]{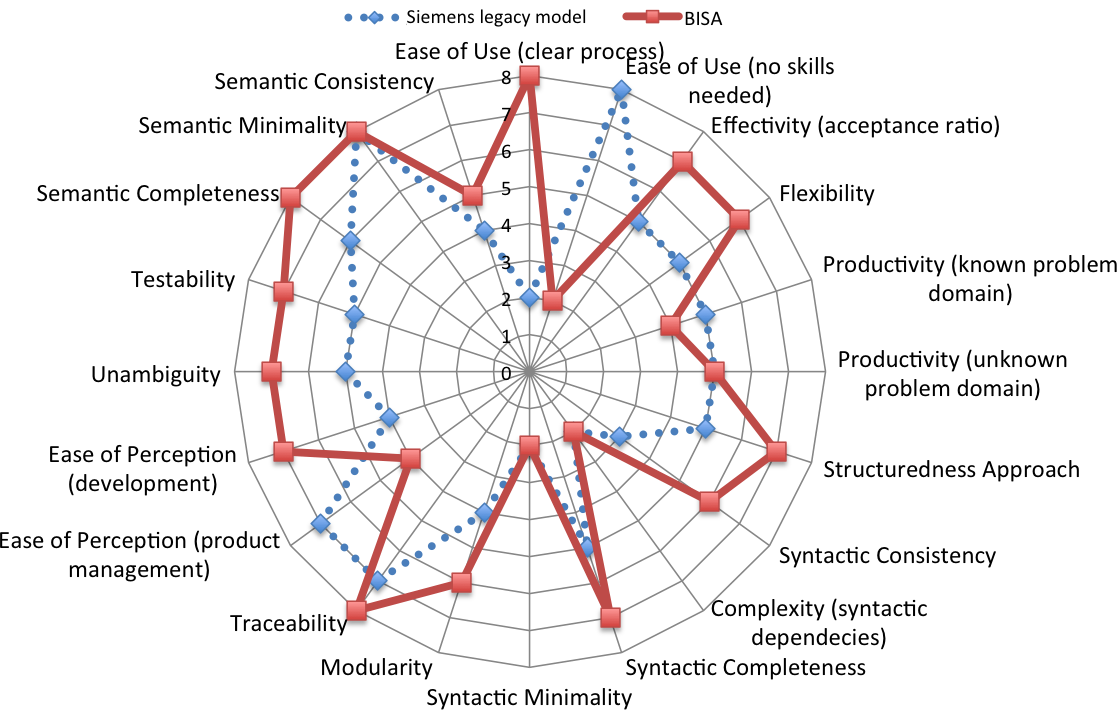}
  }
  
  \subfigure[External rating]{
	\label{fig:siemens2}
	\includegraphics[width=0.6\columnwidth]{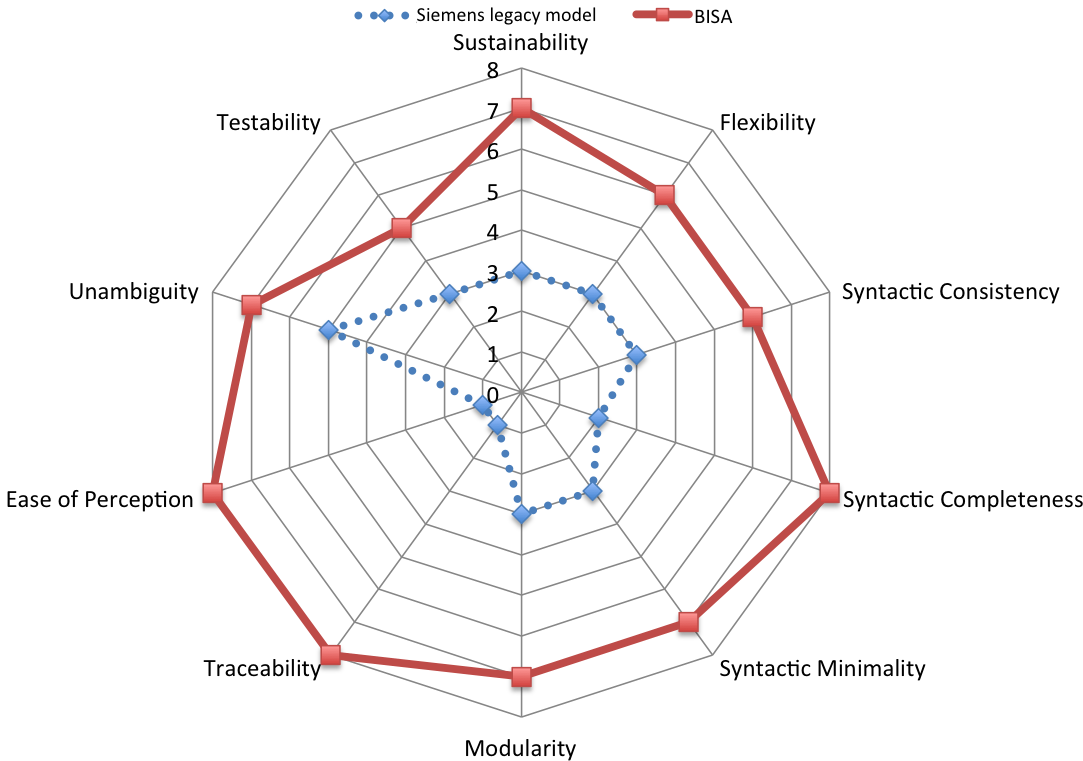}
  }
  \caption{Results from the case study with Siemens.}
\end{figure}

Figure~\ref{fig:siemens1} shows the results from the internal rating at Siemens. We complemented the internal rating with an external one where we called in a neutral person not involved in the RE workshops to support a more unbiased evaluation. The results of the external rating are shown in Fig.~\ref{fig:siemens2}. Both evaluations consider the process when applying the artefact-based RE approach as well as the quality in the created artefacts. Although some criteria had a low consistency between the internal and external rating (e.g., the traceability), the application of the detailed artefact model resulted in general in a better support of creating syntactically consistent and testable results in direct comparison to the previously used RE reference model. The ease of use, however, was rated worse. Details on the the full results can be taken from~\cite{MLPW11}.

The replication studies at Cassidian and BMW revealed that the detailed concept model used to establish the tool infrastructure (see Table~\ref{tab:ArtefactCharacteristics}) again supported the syntactic consistency and completeness in the created artefacts. For both case study systems, we received positive feedback on the artefact-based approach and the evaluation results showed, in contrast to the study at Siemens, improvement in various aspects in the process for creating the artefacts, most importantly, the effectivity and ease of use. Our explanation for the different rating is that the model developed at Capgemini TS has a far more comprehensive concept model capturing all particularities of the appellation domain, which, in turn, implies a stronger learning curve. 

\begin{figure}[htbp]
  \centering
  \subfigure[Applicability evaluation at BMW]{
	\label{fig:bmw1}
	\includegraphics[width=0.47\columnwidth]{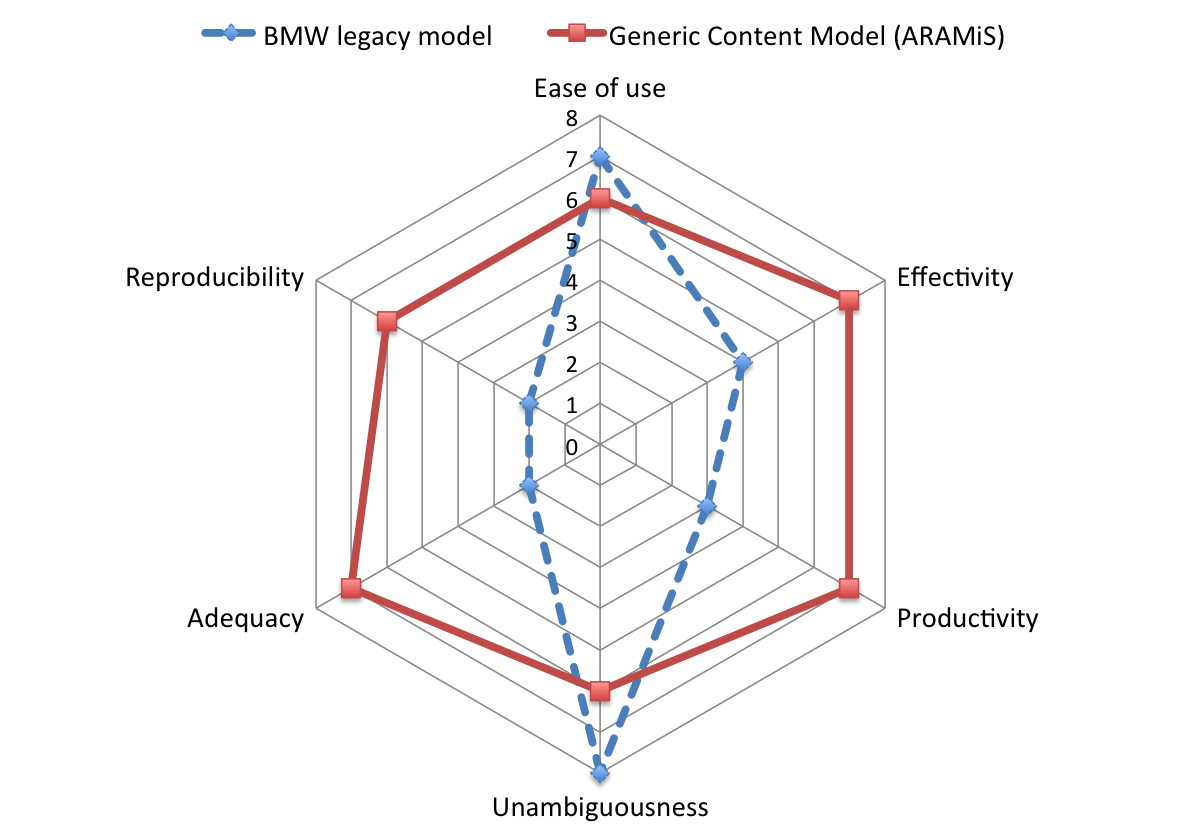}
  }
  \subfigure[Artefact evaluation at BMW]{
	\label{fig:bmw2}
	\includegraphics[width=0.47\columnwidth]{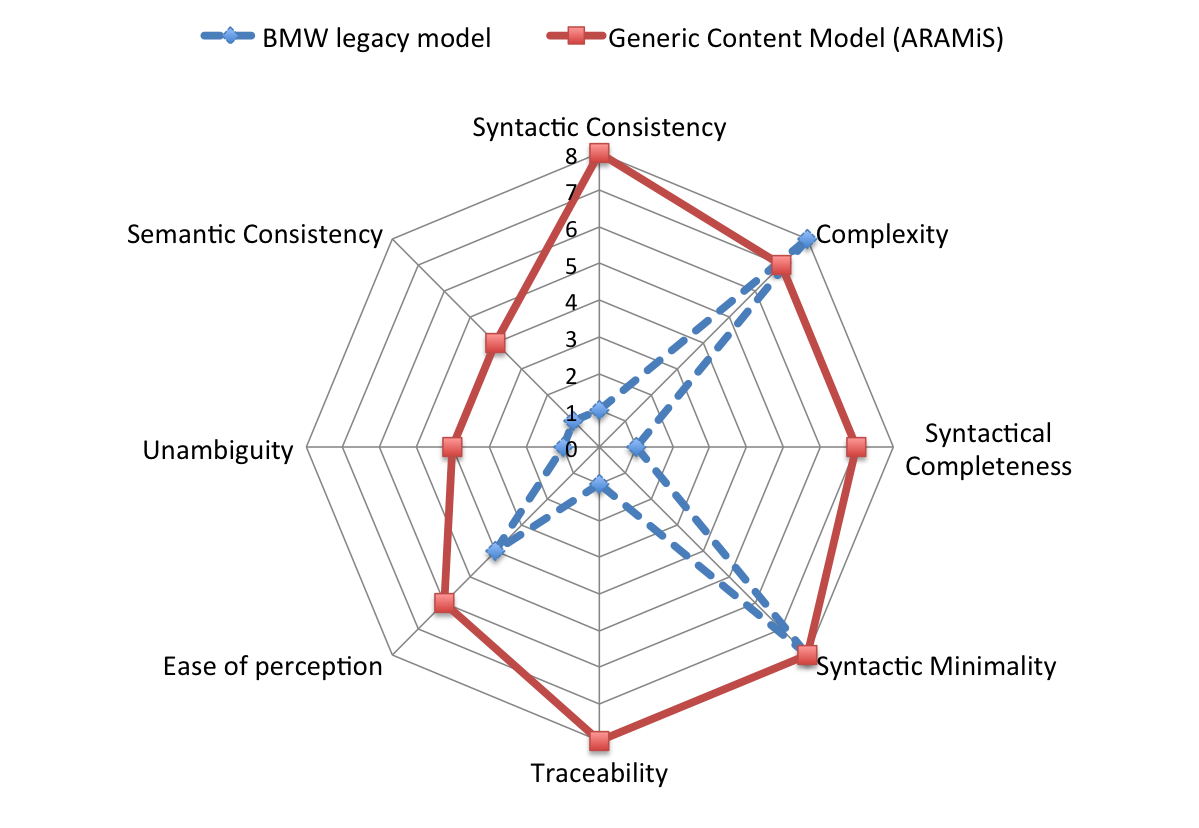}
  }
  \subfigure[Applicability evaluation at Cassidian]{
	\label{fig:cassidian1}
	\includegraphics[width=0.47\columnwidth]{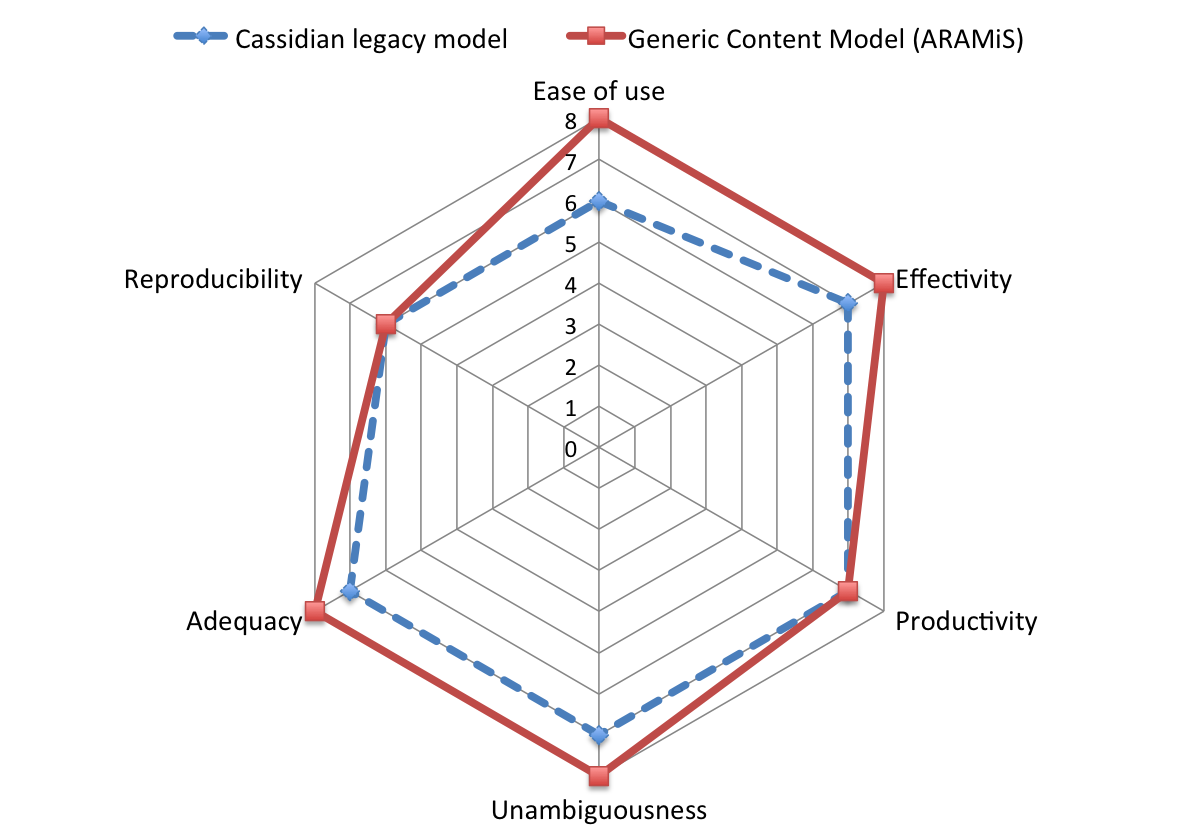}
  }
  \subfigure[Artefact evaluation at Cassidian]{
	\label{fig:cassidian2}
	\includegraphics[width=0.47\columnwidth]{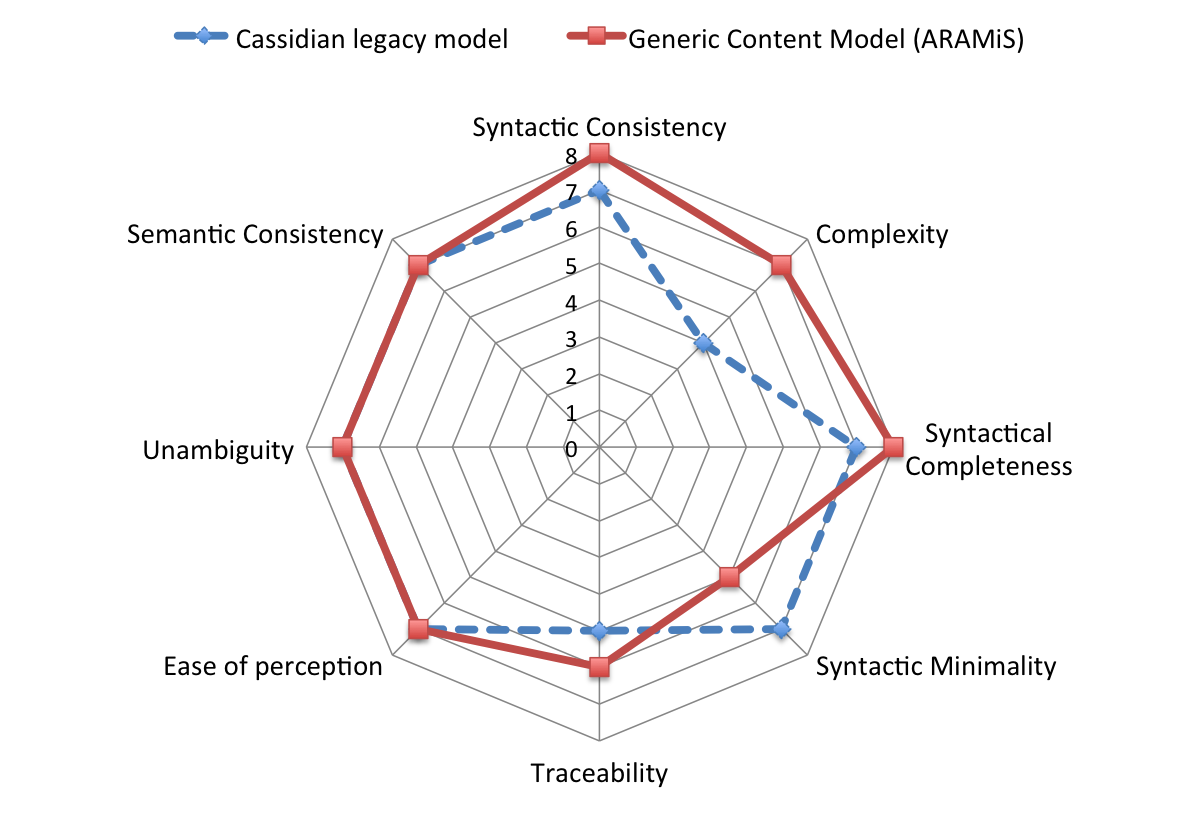}
  }
  \caption{Results from the ARAMiS case studies with BMW and Cassidian.}
\end{figure}

An overview of the rating of the closed questions from our automotive partner BMW is depicted in Fig.~\ref{fig:bmw1} and \ref{fig:bmw2} and from our avionics partner Cassidian in Fig.~\ref{fig:cassidian1} and \ref{fig:cassidian2}. For reasons of illustration, the centre of the Kiviat diagram is labelled again with the value of zero instead of one, since otherwise the data points would overlap in the center.

Both partners seem to perceive the ARAMiS artefact model as improvement in comparison to their previous reference model. The business units at both companies have also incorporated the artefact model into their standard. On first glance, the rating was more critical for both their own and the ARAMiS model by the automotive partner. When analysing the answers to the open questions (the rationale for their ratings), we found that some of the reasons given for a specific rating were quite similar, but rated rather differently. For example, for traceability, both stated that their own reference model provided the possibility to link requirements, but did not provide the possibility to document a rationale. At the same time, their rating of 1 at BMW versus the higher rating of 5 at Cassidian for the criterion \emph{Traceability} shows a considerable difference in perceiving the importance of an artefact-based reference model providing the possibility to document a rationale.

Details on the design and the results of the case studies conducted at Cassidian and BMW can be taken from~\cite{PEM13, PE2012}.

\subsubsection{AMDiRE Case Study}\label{sec:amdirecasestudy}

Apart from the previously conducted and published case studies where we evaluated the models on which AMDiRE relies, we conducted one industrial study where we directly made use of AMDiRE. In contrast to the previously conducted comparative case studies, however, we used AMDiRE as a reference model to build a company-specific artefact model following the procedure illustrated in Sect.~\ref{sec:ExperiencesConstruction} and published in detail in~\cite{MWb13}. 

The case study took place at Wacker Chemie, a company working in the chemical business with quarters in Munich. The department with which we worked focuses on the engineering of development processes for standard as well as custom software development in the company-specific operation processes and their production sites. The goal during the development of the artefact-based RE reference model was to support traceability and testability of the RE artefacts. After a series of analysis workshops and workshops to construct the artefact model (see~\cite{MWb13}), we prepare the evaluation in three pilot projects, which considered internal developments in the company. For reasons of confidentiality, we omit details of the projects and the involved subjects and remain on an abstract description. As a preparation of the evaluation in the pilot projects, we created a short presentation as internal training material and document templates for the new artefacts to be applied in the pilot projects.

The projects covered both standard software development for SAP and custom software development. As project participants served in total 8 employees of which for each project one project lead and one developer were complemented by two process engineers involved in the development of the artefact-based RE reference model. Those latter two process engineers served as coaches to train in a one-day workshop the people for applying the new artefact-based RE reference model developed on basis of AMDiRE. The study design was the same as the one from previously conducted case studies. That is, we conducted workshops to create requirements specifications following the artefact-based RE approach and concluded with an assessment where the project participants rated the new RE approach in direct comparison to the approach previously used in same environment (for previous releases). For the assessment, the project participants were provided again a questionnaire. This questionnaire was similar to the one we used in the previous studies, with minor modifications in wording. For each criteria we used in the rating, we asked a closed question where the participants should rate their agreement to a statement on a Likert-scale from 0 to 6 followed by an open question where the participant could give a rationale for their rating. Table~\ref{tab:questionnaire} gives a condensed view on the closed questions.

\begin{table*}[!tp]
\centering
\caption{Questionnaire for the Assessment (condensed)}
\label{tab:questionnaire}\footnotesize
\begin{tabular}{p{0.3\linewidth}p{0.6\linewidth}} 
\hline
\textbf{Criteria} & \textbf{Statement} \\ \hline 
Flexibility & The RE reference model allows for flexibility. \\
Ease of use & The RE reference model is easy to understand. \\
Effectivity & The RE reference model leads to the desired results.  \\
Efficiency & I perceived the efficiency in the process as high.\\
Customisation & The RE reference model is tailorable according to project-specific situations of the company.\\ 
Process Integration & The RE reference model is integrated into further development activities (e.g., testing) and within the line organisation. \\ 
Structuredness Artefacts & The specification documents are well-suited to be understood respecting their structure by people not being involved into the elaboration of the specifications \\
Syntactic Artefact Quality & The RE reference model supports a high syntactic quality in the created RE artefacts w.r.t. consistency and completeness. \\
Traceability in Artefact & The RE reference model supports traceability within RE (e.g., through rationales) and between RE and further disciplines. \\
Semantic Artefact Quality & The RE reference model supports semantically consistent and complete results. \\
Testability of Artefacts & The RE reference model supports testable RE artefacts. \\ \hline \hline

\end{tabular}
\end{table*}

In the following, we summarise the case study results followed by informal feedback given during the evaluation.

\paragraph{Case Study Results}

\begin{figure}[htbp]
\begin{center}  
\includegraphics[width=1\textwidth]{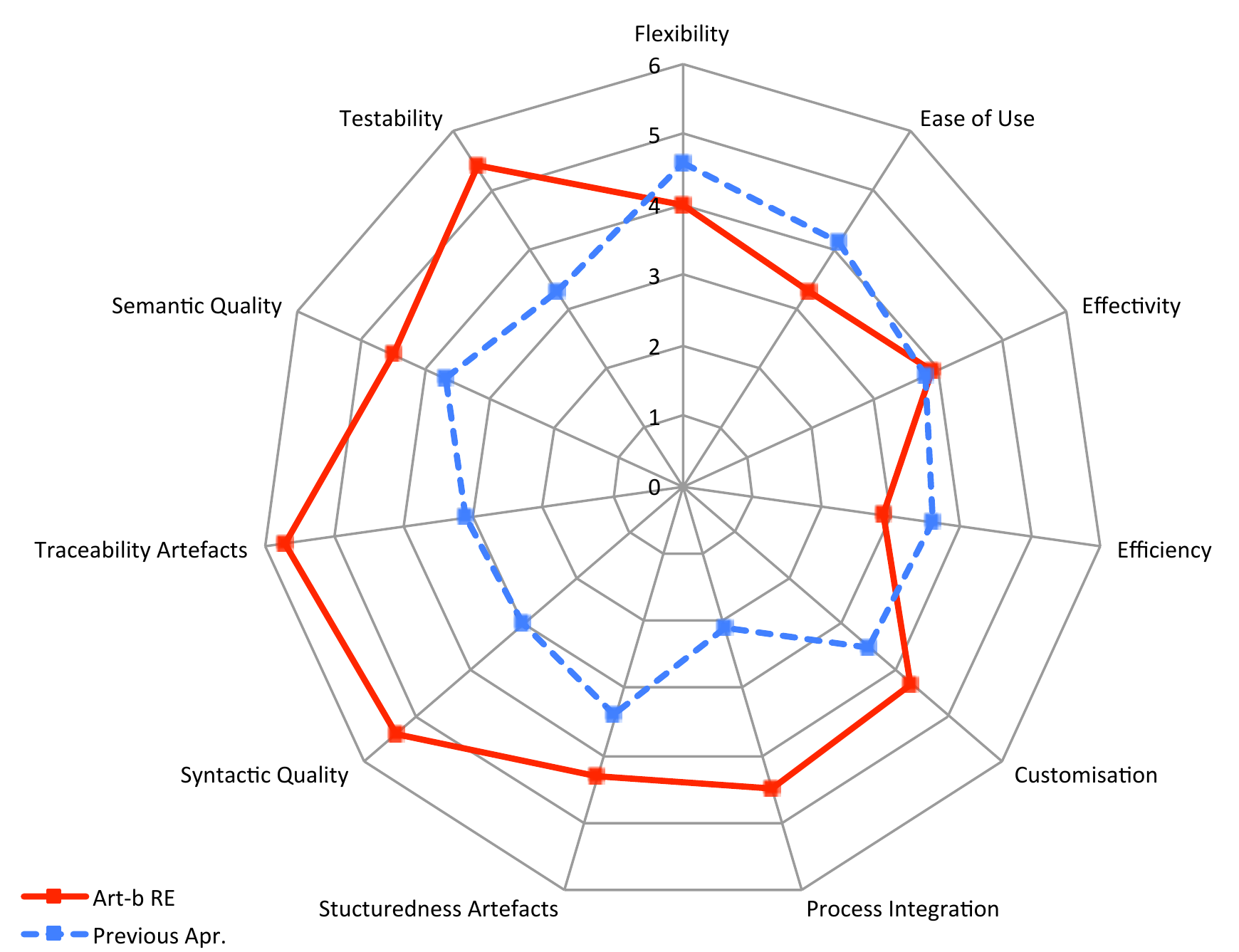}\\
  \caption{Results from case study at Wacker.}
  \label{fig.Wackerstudy}
\end{center}
\end{figure}

Figure~\ref{fig.Wackerstudy} summarises the results from the case study and shows with a solid red line the rating given to the artefact-based RE reference model and with a dashed blue line the rating given for the previously used reference model where the RE artefacts were underrepresented.

The goals of improving the support of traceability and testability were reached and the participants positively rated also other aspects in the artefacts created according to the new reference model. For instance, they saw an increased structuredness in the RE artefacts and rated the process integration to be better than in the previous model as the artefact-based nature clearly defined, for example, roles and responsibilities. 

However, the ease of use was slightly worse than in the previous model, where the feedback indicated the higher level of detail in the reference model to be the main reason. Further feedback that supported this was that participants needed more detailed training as well as comprehensive tool support (going beyond the document templates) to better support the learning curve in applying the new reference model.

\paragraph{Discussion}

Our case study at Wacker supports our claimed general benefits of the artefact-based approach. One remarkable feedback provided by the project participants was, however, the strong need to support the application of the model with tool environments, and detailed training material as, otherwise, the ease of use in the model is strongly hampered. This observation is in tune with the previously conducted case studies where the ARAMiS model resulted to be better rated than the BISA model due to:
\begin{compactenum}
\item providing a less complex structure model that does not focus too much on particularities of single application domains, while
\item hiding the remaining complexity of the underlying concept model with its a directly applicable tool environment.
\end{compactenum}

Although the AMDiRE approach covers both characteristics, we have now reached the point where we need to conduct external evaluations by unbiased researchers and practitioners not involved in the development of AMDiRE (see also the research method in the introduction on page~\pageref{sec:researchmethod}). This will also show to what extent AMDiRE can be used by others and what exact effects this will have on aspects negatively rated at Wacker (relating to the learning curve).

To this end, we make publicly accessible the tools, the models, and the examples as well as the evaluation templates related to AMDiRE~\cite{onlineAMDiRE} and encourage researchers and practitioners to critically discuss and evaluate our approach in external contexts. 

The article at hand thus builds the first step in the dissemination of our research results not only into practical environments with isolated problems and goals, but also back into the research community.

\subsection{Lessons Learnt}
Throughout the past six years of collaborating with industrial partners on improving their requirements engineering practices, we have collected a number of lessons learnt, which we summarise in the following.

\paragraph{Tradeoff between high level of detail and usability.} Regarding more general lessons learnt in the field of artefact orientation, we have observed more detailed artefact models to increase the quality of the results as they can give detailed guidance on the concepts to use when specifying the contents. At the same time, however, those more detailed models constrain the ease of use as they implicate a higher learning curve. Obviously, simpler artefact models have the opposite effect. Whereas they are easy to apply and understand, they cannot give detailed guidance on creating detailed, syntactically consistent contents. The only mitigation we see to this problem so far is to explicitly point out this tradeoff as early as possible and in case of focussing on detailed concept models providing tool support right from the beginning.

\paragraph{Terminology.} In every workshop we have held, the longest time was usually spent on terminology discussions. Either there were different terms available and the discussion was which one of them should be given preference, for example \emph{context} versus \emph{environment}. Or there was one term but different stakeholders had different interpretations of the concept it represented, for example \emph{function}. These discussions are time-consuming but unavoidable and crucial to make sure the artefact model will be accepted by all stakeholders later on. The mitigation strategy we followed was to explicitly reserve a time slot early on to establish a list of most important terms and definitions that have to be agreed on.

\paragraph{Maintainability.} Furthermore, we noticed the maintenance of established artefact models to be challenging. Once detailed models are established, it is difficult to integrate new concepts due to the complex content-related dependencies. We are not aware of any real mitigation to this point, but we suggest that every single modification of the model should be justified, carefully planned, well documented, and explicitly evaluated in an own pilot project to detect potential inconsistencies arising from the modification in the model as early as possible. 

\paragraph{Incorporate down-to-earth informality.} The biggest lesson learnt, although not surprising in itself but only in its extent, is the down-to-earth informality that is residing in the daily business of software systems development. This is in contrast to the high quality requirements and constraints that have to be adhered to by the final products. There is a high demand for easy to use approaches. The more details a concept model has, the more should be invested in pilot studies, coaching material, and guidelines (reflected, e.g., in the effort spent for the development of the approach at Capgemini TS). 

\paragraph{Support learning curves.} One of the quality characteristics that has always been rated worse for the new artefact-based approaches compared to the previously residing approaches is \emph{efficiency}. Our assumption is that this is not only due to the higher simplicity of the previous models but also that the previous approaches are what the developers are accustomed to. Consequently, anything that is different from what they are used to will initially require a slightly higher effort and might therefore easily leave the impression of decreased efficiency. As a matter of fact, this strongly relates to a learning curve that is always implied by new approaches~\cite{Spence1981}. We encountered two means to support learning curves: continuous pilot studies and training and coaching. The first includes a smooth integration of developed approaches via pilot phases, which we use to get additional feedback from the project environments, and extends to long-term studies to investigate the benefits of improvement endeavours w.r.t. to project quality. The latter includes the preparation of guidance, which we believe to build an essential building block for not only introducing a new approach but for establishing it in a long run taking into account the different organisational cultures. Unfortunately, the writing of a handbook or the development of course material and trainings is often neglected. In that case the developers are likely to be resistant against innovations.

\paragraph{Tackle distrust right from the beginning.} In projects, we have often encountered a certain distrust against new approaches. The willingness to change is rather low and a commonly known problem~\cite{MW13}, i.e., the willingness to change results from the initial effort required to even try a different way of working, but also from general beliefs, experiences, expectations, emotions and desires. There is no universal silver-bullet to tackle this problem as, more generally speaking, the success of any improvement endeavour also depends on social skills~\cite[chp.~10]{everett95} and eventually on politics~\cite{MM11}. We experienced, however, the deep involvement of different stakeholders (especially from project settings) right from the beginning of an improvement project and an honest communication are the most important aspects to tackle distrust (see also~\cite{MWb13}).

\section{Conclusion}
\label{sec:Conclusion}

This article described the fundamentals in artefact-based requirements engineering and presented the AMDiRE approach that emerged from six years of experience with developing artefact models and integrating them into the surrounding requirements engineering processes of various socio-economic contexts. Furthermore, we presented our evaluations in different industrial case studies and their replications in different companies as well as lessons learnt from these collaborations.

A subset of our case studies were performed as comparative studies that evaluate the application of the artefact-based RE approaches in direct comparison to the activity-based approaches previously used in the same environment. We could show how artefact orientation supports for detailed, consistent results while supporting the flexibility in the process which is especially important to RE. We also discussed, however, that we continuously need to make a trade-off between a detailed content model to support high quality results, and the resulting higher learning curve that affects the ease of use of artefact-based RE. 

Our work has been incorporated into the daily requirements engineering practices of the companies with whom we collaborated. We strongly believe that this approach of carrying research into practice, adapting it to the business context, and establishing it in collaboration with the practitioners is a sustainable way of promoting the use of current research results. We hope to carry this work further into the requirements engineering community to encourage the application of artefact-based approaches and to give an example of an artefact model that can be used in a variety of application domains. 

Researchers can already build their fundamental, educational, and evidence-based work upon our contributions. Our empirical results already give the opportunity to steer their requirements engineering research in a problem-driven manner. Practitioners can furthermore directly apply our model in their own socio-economic contexts with the awareness of the benefits and shortcomings of the incorporated concepts. We thus laid the first fundamental, conceptual, and empirical basis for artefact-based RE research. However, we do not claim that our approach suits every situation and that our empirical results allow for perpetually valid generalisations as we always focused with our action research case studies on specific socio-economic contexts. In fact, the research area needs further investigation and the case studies need further independent replications. The reason for independent replications is that if we go to another company and help them improve their requirements engineering using our model (which is the only reason for a company to hire us), this probably biases the results towards our model as companies get free consulting during the process. Therefore, we need an external third party to perform the validation. 
Consequently, we encourage researchers and practitioners to critically discuss our approach and to join us in the empirical evaluations of artefact-based RE.

\paragraph{Future Work.} We are encouraging further replications of our comparative studies in different application domains and development contexts. 
Another objective for future work is to further facilitate the dissemination and empirical evaluation of our approach. We have developed a first prototype for tool support as a profile with a respective template in MagicDraw that is freely available~\cite{onlineAMDiRE}. We use our tools as an additional means for the dissemination and for continuous evaluations in pilot projects to gather more information and user feedback while preparing the development of a more elaborated stand-alone tool.


\newpage
\appendix
\section{AMDiRE Content Model}
\label{sec:ContentModel}

The following appendix defines the content model of AMDiRE in detail giving for each content item a definition of the used concepts.

The artefact model is specified using the following notational aspects of UML class diagrams:
\begin{itemize}
\item We denote the hierarchical structuring of the structure model with packages.
\item For the definition of the content model, we use a class diagram.
\item For content items that are crucial for only a specific application domain, but irrelevant for another, we use the stereotype \emph{$<<$Domain$>>$}, such as business process models being crucial for the domain of business information systems, but irrelevant for the domain of embedded reactive systems.
\end{itemize}

\subsection{Context Specification}
\begin{figure}[!h]
\begin{center}  
\includegraphics[width=1\textwidth]{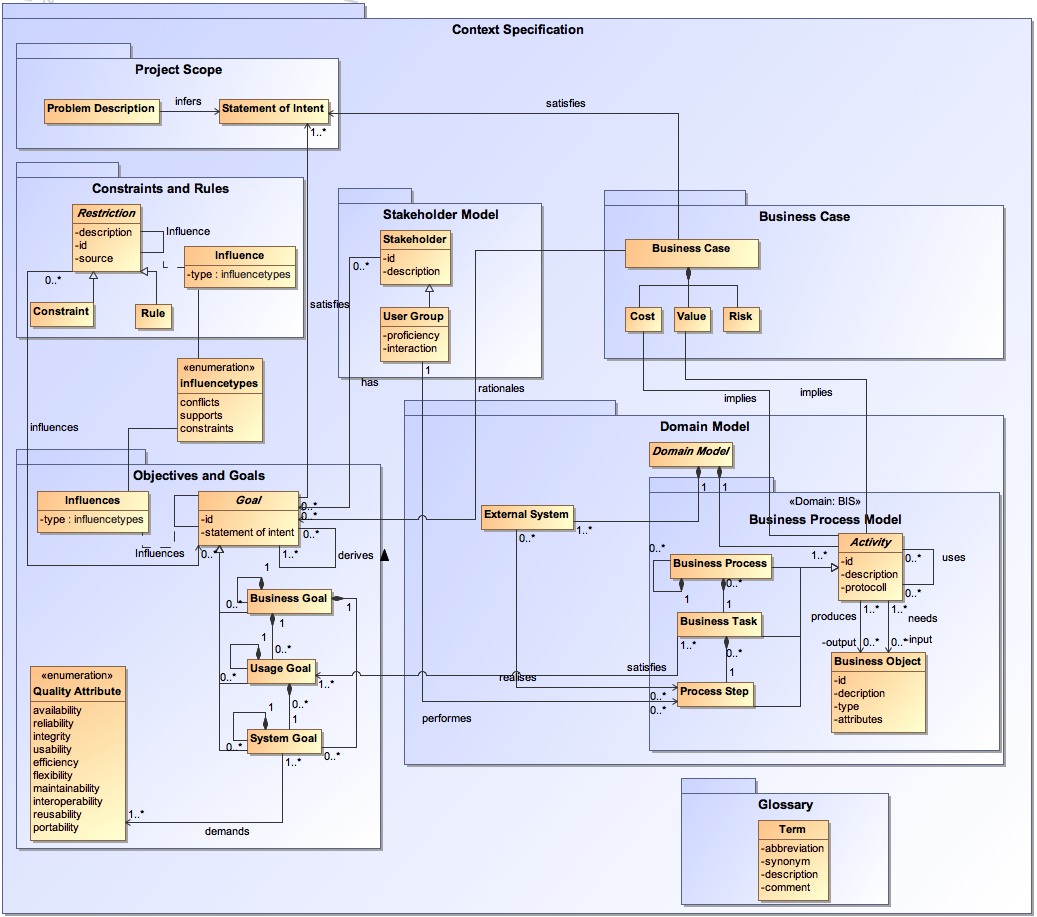}\\
  \caption{The AMDiRE context specification.}\label{fig.amdire_context}
\end{center}
\end{figure}

The context specification is depicted in Fig.~\ref{fig.amdire_context}. It contains the Project Scope, the Constraints and Rules, the Stakeholder Model, the Business Case, the Objectives and Goals, the Domain Model, and the Glossary.

\newpage

A description of the content items is provided in Table~\ref{tab:CIContextSpecification}.

\footnotesize
\begin{longtable}{p{0.15\linewidth}p{0.12\linewidth}p{0.65\linewidth}}
\caption{Content items in the Context Specification.}\\
\label{tab:CIContextSpecification}\\
\hline
\textbf{Content Item} & \textbf{Exempl. Notation} & \textbf{Description} \\  
\hline
\hline
Project Scope & Natural text & Content item of the \emph{Context Specification} that consists of a \emph{Problem Description} and a \emph{Statement of Intent}, i.e.\ a conclusion of the objectives of a potentially resulting project. \\

Constraints \newline and Rules & \mbox{Natural text}, graphs  & Restrictions that can influence each other either in form of support or a conflict. We distinguish \emph{Constraints} as not negotiable restrictions in the domain and \emph{Rule}, often referred as conditional standard procedures~\cite{BABOK09}\\

Stakeholder Model & \mbox{UML actor} hierarchy, tables & \emph{Stakeholders} comprehend individuals, groups, or institutions having the responsibility for requirements and a major interest in the project~\cite{Cleland05}, while \emph{User Groups} are a specialisation of stakeholders with a particular proficiency and involvement in interaction with the system~\cite{FODA}. \\

Business Case & Natural text & Described and detailed using the additional elements of Cost, Value, and Risk. The Business Case satisfies the Statement of Intent from the Project Scope and rationalises the goals in the content item Objectives and Goals \\

Objectives \newline and Goals &\mbox{Goal graphs} (e.g.\ KAOS) & Each Goal, whether it is a \emph{Business Goal}, a \emph{Usage Goal}, or a \emph{System Goal}, is issued by a \emph{Stakeholder}. Goals satisfy the \emph{Statement of Intent}~\cite{Lams09}, they build a hierarchy, and they can influence each other in terms of conflicts, constraints, or support. Each usage goal is related to a business goal and each system goal to a usage goal. Furthermore, system goals demand one or more \emph{Quality Attributes}~\cite{ISO25030}. \\

Domain Model & \mbox{UML activity} \mbox{diagrams or} BMPN & The External Systems, that interact with the system under development, compose the \emph{Domain Model}. For business information systems, the domain model is extended with a \emph{Business Process Model} represented in various types of \emph{Activities} that need and produce \emph{Business Objects}. A business process model is a collection of all instances of the activities and their (causal) relations, performed by roles in order to produce some outcome of value~\cite{LK06}. The activities can be defined as an abstract \emph{Business Process}, a \emph{Business Task} (business use case), or an atomic \emph{Process Step} -- the latter represents single atomic steps performed by a \emph{User Group}. \\

Glossary & Structured text &This content item contains all important \emph{Terms} for the system under consideration and the respective project management, including their abbreviation, synonyms, and description. It shows up as well in Requirements Specification and the System Specification as more Terms are added over the course of the project.\\
\hline
\end{longtable}
\normalsize
\newpage

\subsection{Requirements Specification}

\begin{figure}[!h]
\begin{center}  
\includegraphics[width=1\textwidth]{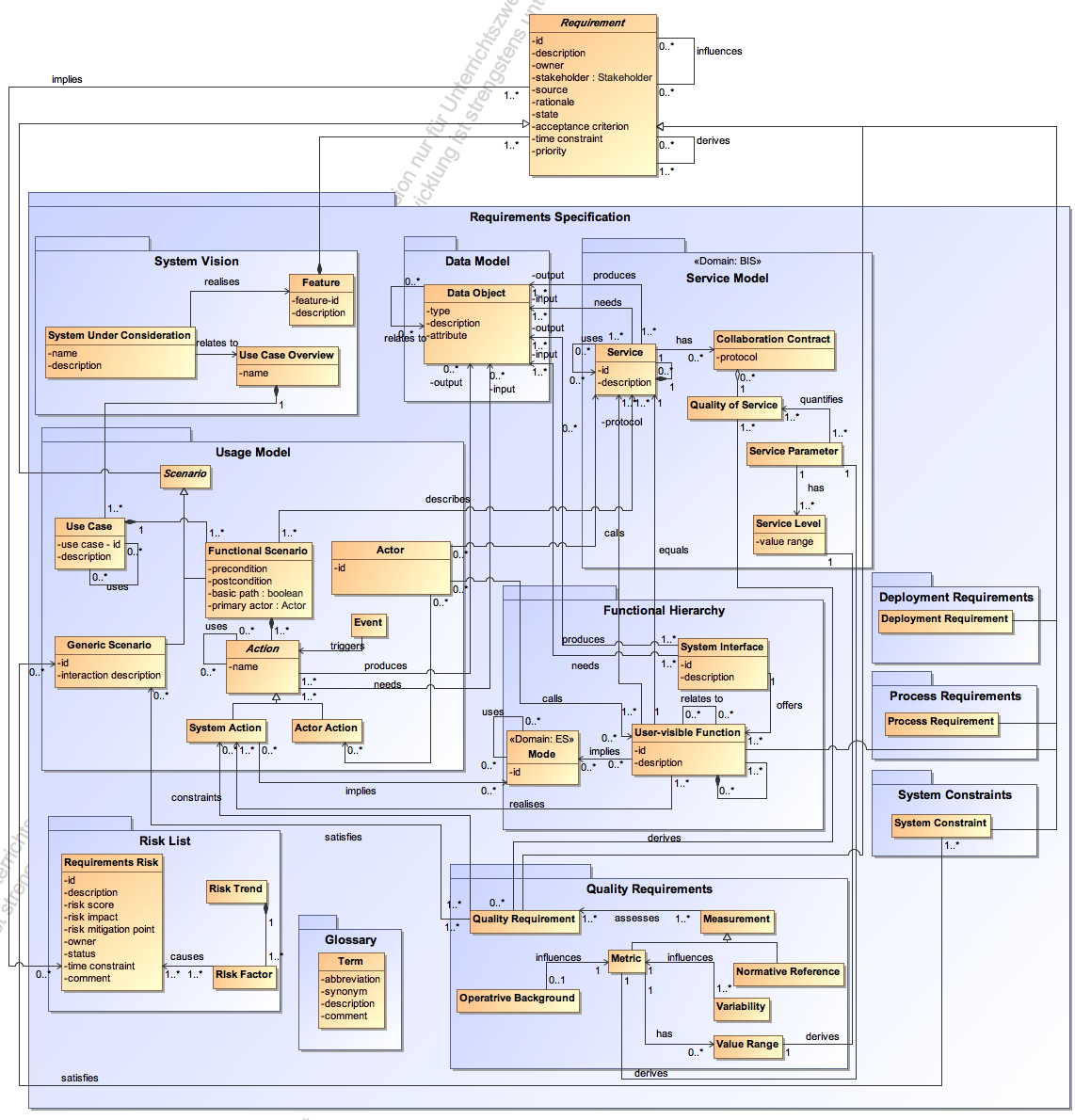}\\
  \caption{The AMDiRE requirements specification.}\label{fig.amdire_requirements}
\end{center}
\end{figure}

The requirements specification is depicted in Fig.~\ref{fig.amdire_requirements}.
It contains the System Vision, the Usage Model, the Data Model, the Service Model, the Functional Hierarchy, the Quality Requirements, the Deployment Requirements, The System Constraints, the Process Requirements, and the Risk List. A description of the content items is provided in Table~\ref{tab:CIRequirementsSpecification}.
\newpage

\footnotesize
\begin{longtable}{p{0.15\linewidth}p{0.12\linewidth}p{0.65\linewidth}}
\caption{Content items in the Requirements Specification.}\\
\label{tab:CIRequirementsSpecification}\\
\hline
\textbf{Content Item} & \textbf{Exempl. Notation} & \textbf{Description} \\  
\hline
\hline
System Vision & Rich picture & The system vision comprehends the system context of the \emph{System Under Consideration}, which is intended to realise a number of \emph{Features}. A feature is, in our understanding, a prominent or distinctive user-recognisable aspects, quality, or characteristics of a system that is related to a specific set of requirements, whose realisation enable the feature~\cite{CL08, FODA}. In addition to features, we specify an \emph{Use Case Overview}, i.e.\ a (potentially) graphical overview of the use cases specified in full in the usage model.
\\
Usage Model & Structured text, \mbox{UML activity} diagrams & This content item details the \emph{Use Case Overview} of the \emph{System Vision} in its \emph{Use Cases}. We distinguish \emph{Services} and \emph{Use Cases}. Both concepts are means to describe (black box) system behaviour. \emph{Use Cases} describe sequences of interaction between \emph{Actors} (\emph{realising} user groups) and the system as a whole. More precisely, a use case represents a collection of interaction scenarios, each defining a set of interrelated actions that either are executed by an actor or by the system under consideration~\cite{Cock00}. For each use case, there is at least one \emph{Functional Scenario} in which \emph{Actors} participate. A Scenario inherits from a requirement (not a whole use case) and each Scenario is detailed into Actions, which can be \emph{Actor Actions} or \emph{System Actions} each processing \emph{Data Objects}. Functional scenarios are triggered by Events. Furthermore, we include \emph{Generic Scenarios}, which serve for the satisfaction of \emph{Quality Requirements} as they provide a means to specify generic interactions between actors and a system not necessarily motivated by business processes, such as maintenance activities an administrator performs.
\\
Service Model & Graphs & This content item is relevant for the domain of business information systems and specifies \emph{Services}  with a \emph{Collaboration Contract} and a defined \emph{Quality of Service}. \emph{Services} describe a logical representation of a use case, not necessarily involving actors or concrete sequences of interaction. A service represents user-visible functions that the system shall offer and is described via input/output-relations~\cite{HT09, Rittmann08}, i.e., the mapping of (typed) inputs and outputs, both represented as information system objects. The quality of service is described using \emph{Service Parameters} (corresponding to \emph{Metrics} from the \emph{Quality Requirements}) that have particular \emph{Service Levels}. If a collaboration contract specifies a set of service (calls) as part of a causal relation, we use scenarios within the \emph{Functional Scenarios} of the \emph{Usage Model}.
\\
Data Model & mbox{UML class} \mbox{diagrams} & The \emph{Data Model} contains all objects processed as part of functions and interaction scenarios.
\\
Functional \newline Hierarchy & Graphs, \mbox{I/O-tables} & The \emph{User-visible Functions} equal the \emph{Services} from the BIS-specific \emph{Service Model} and realise the \emph{System Actions} from the \emph{Usage Model}. Functions are organised in a hierarchy to build the transition to the system specification as they describe a logical representation of a system action and offered by typed interfaces. When a Function is triggered, we can optionally define \emph{Modes}, which we use in the \emph{System Specification} for the definition of detailed \emph{Behaviour Models}. \emph{Services}, in turn, describe a complementary logical representation of a use case, not necessarily involving actors or concrete sequences of interaction. They furthermore offer \emph{Interfaces}, which are typed according to the \emph{Data Objects} of the \emph{Data Model}.\\
Quality \newline Requirements &Natural text & Quality requirements are assessed by \emph{Measurements} that can be either a \emph{Normative Reference} (e.g.\ a GUI style guide) or a \emph{Metric}. Quality Requirements constrain \emph{System Actions} and can be satisfied by \emph{Generic Scenarios}. We make use of quality definition models as by Deissenb\"ock et al.~\cite{DJLW09}.
\\
Deployment \newline Requirements & Natural text & \emph{Deployment Requirements} describe demands towards the deployment procedure, constraining the process design of the deployment, and the technical environment during initial launch of the system or specific parts of it. \\
System \newline Constraints & Natural text & The system's constraints describe logical and technical restrictions on a system's architecture, its functionality by means of single atomic actions, and its quality by means of assessable system quality requirements. We consider concepts that describe the transition to logical and technical architecture layers acc. to~\cite{Wieg03}. Hence, we see a system as a grey box rather than as a glass box, since we restrict systems' internals, but do not consider their logical structure by interacting components, interface specifications, and functions. \\
Process \newline Requirements & Natural text & \emph{Process Requirements} constrain the content and / or structure of selected artefact types and the process model, i.e., the definition of the milestones regarding time schedules, used infrastructure like mandatory tools, and compliance to selected standards and software process models. \\
Risk List & Natural text & The \emph{Risk List} includes a description of all risks that are related to project-specific requirements. 
The conceptualisation of requirements risks is considered on the basis of an artefact model~\cite{Isl09, IHMJ09}. The \emph{Requirements Risks} are implied by the various types of Requirements and we use the risk list as an interface to risk management. Each risk is caused by a \emph{Risk Factor}. A \emph{Risk Trend} is composed by all Risk Factors.\\
\hline
\end{longtable}
\normalsize

\newpage

\subsection{System Specification}

\begin{figure}[!h]
\begin{center}  
\includegraphics[width=0.75\textwidth]{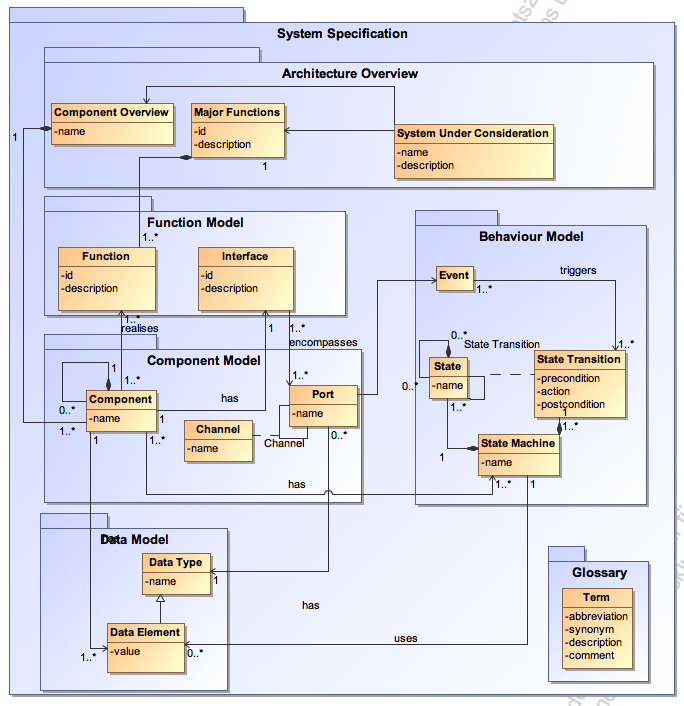}\\
  \caption{The AMDiRE system specification.}\label{fig.amdire_system}
\end{center}
\end{figure}

The system specification is depicted in Fig.~\ref{fig.amdire_system}.
It contains the Function Model, the Component Model, the Behaviour Model, the State Model, and the Data Model. A description of the content items is provided in Table~\ref{tab:CISystemSpecification}.

\footnotesize
\begin{longtable}{p{0.15\linewidth}p{0.12\linewidth}p{0.65\linewidth}}
\caption{Content items in the System Specification.}\\
\label{tab:CISystemSpecification}\\
\hline
\textbf{Content Item} & \textbf{Exempl. Notation} & \textbf{Description} \\  
\hline
\hline

Architecture Overview & Component \mbox{diagram} & The \emph{Architecture Overview} includes the \emph{Component Overview} of the \emph{Components} as well as the \emph{Major Functions} that summarize the \emph{Functional Hierarchy}.\\
Function Model & Graphs, tables & The \emph{System Functions} offered by the \emph{System Interface} realise the \emph{User-visible Functions} from the \emph{Functional Hierarchy}. The system interface encompasses the Ports detailed in the Component Model.\\
Component Model & Component \mbox{diagram} & The component model describes the logical component architecture in detail including \emph{Ports} (building typed interfaces) and \emph{Channels} over which components communicate. Components can be decomposed into more further sub-components~\cite{TUM-I0816}. \\
Behaviour Model & Automata & The behaviour of components is specified with by \emph{Events}, \emph{States} and \emph{State Transitions}. The resulting state machines specify the behaviour and the used data elements of the data model. \\
Data Model & \mbox{UML class} diagram, \mbox{dictionary} & We specify data elements of a certain type and their relations as they result from the behaviour model. To this end, \emph{Data Elements} refine \emph{Data Objects} from the requirements specification and have a particular \emph{Data Type}.\\
\hline
\end{longtable}
\normalsize

\newpage

\bibliographystyle{spbasic}
\bibliography{amdire}
\end{document}